\documentclass[aps,prd,preprint,nofootinbib]{revtex4-1}

\usepackage{graphicx}
\usepackage{amsmath}
\usepackage{mathtools}
\usepackage{amssymb}
\usepackage{hyperref}
\usepackage{color}
\usepackage{cleveref}
\usepackage{comment}
\usepackage{dsfont}

\crefname{figure}{Fig.}{Fig.}
\Crefname{figure}{Figure}{Figures}
\crefname{equation}{Eq.}{Eq.}
\Crefname{equation}{Equation}{Equations}
\crefname{section}{Sec.}{Sec.}
\Crefname{section}{Section}{Sections}
\crefname{appendix}{App.}{App.}
\Crefname{appendix}{Appendix}{Appendices}
\creflabelformat{equation}{#2\textup{#1}#3}

\newcommand{\infinity}{\infty}

\newcommand{\Hess}{\mathrm{Hess}}
\newcommand{\sgn}{\mathrm{sgn}}

\definecolor{darkgreen}{rgb}{0,.7,0}
\definecolor{linkblue}{rgb}{0.,0.,0.9333}

\makeatletter
\newcommand{\pushright}[1]{\ifmeasuring@#1\else\omit\hfill$\displaystyle#1$\fi\ignorespaces}
\newcommand{\pushleft}[1]{\ifmeasuring@#1\else\omit$\displaystyle#1$\hfill\fi\ignorespaces}
\makeatother

\hypersetup{
    pdffitwindow=true,      
    pdfauthor={W. A. Horowitz and A. Rothkopf},     
    bookmarksnumbered=true,
    colorlinks=true,       
    linkcolor=red,          
    citecolor=darkgreen,        
    filecolor=magenta,      
    urlcolor=cyan,           
    menucolor=blue,
    breaklinks=true,
}

\begin{document}
\vspace*{-0.7in}

\title{Hamilton Revised: \\ The Action Principle for Initial Value Problems}

\author{W.\ A.\ Horowitz}%
\email{wa.horowitz@uct.ac.za}
\affiliation{%
 Department of Physics, University of Cape Town, Private Bag X3, Rondebosch 7701, South Africa
}
\affiliation{%
Department of Physics, New Mexico State University, Las Cruces, New Mexico, 88003, USA
}
\affiliation{%
Theoretical Sciences Visiting Program, Okinawa Institute of Science and
Technology Graduate University, Onna, 904-0495, Japan
}
\author{A.\ Rothkopf}%
\email{akrothkopf@korea.ac.kr}
\affiliation{%
 Department of Physics, Korea University, Seoul 02841, Republic of Korea
}

\date{\today}

\begin{abstract}
    We present the variational action principle for initial value problems in classical, conservative-force point particle mechanics. We rigorously derive this formulation by taking the classical limit of the Schwinger-Keldysh expression for the time dependence of the expectation value for operators in quantum mechanics. We clarify the connection between the variation of the position and the variation of the velocity of a particle when implementing Hamilton's Principle in deriving the Euler-Lagrange Equations. We show that both the plus and minus Keldysh paths (of the average and difference of the forward/backward paths) have classical paths and fluctuations---unlike the common perception that the minus path provides the fluctuations around the single classical solution given by the plus path---and that the fluctuations of both paths are crucial for the correct normalization of the classical limit.  The classical limit yields ``initial conditions'' and equations of motion for the minus paths such that the unique classical solution for the minus paths is that they are identically zero, and, fascinatingly, that the minus paths' solution propagates backwards in time; thus one does not need to set the minus paths to zero by hand when taking the classical limit of the Schwinger-Keldysh formalism. We note implications for the classical and quantum mechanics of non-holonomic constraints and quantum field theories with gauges dependent on the derivatives of the fields.
\end{abstract}

\maketitle

\section{Introduction}
The field of classical point particle mechanics concerns itself with the deterministic prediction  of the future position of a particle given a particular physical setup.  The setup will in general set the initial position and velocity of the particle and provide some notion of the forces that will act on the particle.  Classical mechanics is, in principle, fully understood: Newton's Second Law \cite{Newton1687}
\begin{align}
    \sum_i\vec F_i(t) = m\left.\frac{d^2x}{dt^2}\right|_t
\end{align}
gives a second order differential equation in time for the position $x(t)$ of a particle of mass $m$ subject to external forces $\vec F_i(t)$.  Given the initial position $x(t_i)=x_0$ and initial velocity $\left.\frac{dx}{dt}\right|_{t_i}=v_0$, the Picard-Lindel\"of Theorem \cite{CoddingtonLevinson1955} guarantees that a solution of the second order differential equation exists and is unique.  

Despite its well-deserved fame and utility, Newton's Second Law has a number of drawbacks.  Centuries of work led to the understanding that, in many ways, a far better formulation of classical point particle mechanics can be found in terms of Hamilton's Principle of Extremized Action \cite{Hamilton1834,Hamilton1835}, that the path of the classical point particle $x_{cl}(t)$ satisfies
\begin{align}
    \delta S|_{x_{cl}(t)} = 0,
\end{align}
where the action
\begin{align}
    S \equiv \int dt \, L
\end{align}
is defined in terms of the Lagrangian
\begin{align}
    L\equiv T-V,
\end{align}
where $T$ is the kinetic energy and $V$ is the potential energy, and $\delta S$ is the first variation of the action.  Hamilton's formulation has enormous advantages over Newton's \cite{Arnold1989,Goldstein2002}.  A proper subset of these advantages includes
\begin{itemize}
    \item making the role of symmetries transparent via Noether's Theorem \cite{Noether1918}
    \item formulating the problem in terms of scalar, rather than vector, quantities
    \item allowing for the easy use of generalized coordinates
    \item readily incorporating constraints (whose explicit forces do not need to be known \emph{a priori})
    \item easily generalizing to continuous distributions of masses and to fields
    \item leading to a phase space manifold with symplectic structure, providing a ``royal road'' to quantum mechanics
    \item allowing for global rather than local solutions of classical paths \cite{RothkopfHorowitz2024APS}
\end{itemize}

Despite these enormous advantages, as well as $\sim200$ years of study, there are still several conceptual and practical open questions related to Hamilton's formulation.  To understand the conceptual issues, let us briefly review the usual derivation of the Euler-Lagrange equations from Hamilton's Principle \cite{Arnold1989,goldstein2002classical}.  We start with a Lagrangian $L[\vec q(t),\dot{\vec q}(t)]$ that depends on the path of the particle in generalized coordinates $\vec q(t)$ and the time derivative of the generalized coordinates $\dot{\vec q}(t)\equiv \left.\frac{d\vec q}{dt}\right|_t$.  Then the variation of the action yields
\begin{align}
    \delta S = \int_{t_i}^{t_f}dt\, \left[ \frac{\partial L}{\partial q^i}\delta q^i(t) + \frac{\partial L}{\partial \dot q^i}\delta \dot q^i(t) \right].
\end{align}
For a particle undergoing unconstrained motion, one has the transposition rule \cite{Flannery2011DAlembertLagrange}
\begin{align}
    \label{eq:transpositionrule}
    \delta \dot q^i(t) = \frac{d}{dt}\big[ \delta q^i(t)\big].
\end{align}
One then integrates by parts, which yields
\begin{align}
    \delta S = \left. \delta q^i\frac{\partial L}{\partial \dot q^i} \right|_{t_i}^{t_f} + \int_{t_i}^{t_f}dt\,\delta q^i\left[ \frac{\partial L}{\partial q^i} - \frac{d}{dt} \frac{\partial L}{\partial \dot q^i} \right].
\end{align}
The endpoints of the path are taken to be fixed, in which case $\delta q^i(t_i)=\delta q^i(t_f) = 0$, and the boundary terms are identically 0.  Then, because the variations in the path $\delta q^i(t)$ are independent and arbitrary, in order for the variation to be identically zero, we must have by the Fundamental Theorem of the Calculus of Variations \cite{GelfandFomin2000} the Euler-Lagrange Equations,
\begin{align}
    \frac{\partial L}{\partial q^i} - \frac{d}{dt} \frac{\partial L}{\partial \dot q^i} = 0.
\end{align}
One then solves the Euler-Lagrange Equations as an initial value problem, in which $q^i(t_i)=q_0^i$ and $\dot q^i(t_i) = \dot q_0^i$.

There are two main conceptual issues with the above derivation that we wish to highlight.  The first conceptual issue is the imposition of \emph{boundary value problem} conditions on the variation of the path at the initial and final times, $\delta q^i(t_i)=\delta q^i(t_f)=0$ in the derivation of the Euler-Lagrange Equations, which are then posed and solved for the particle's path $q^i(t)$ in terms of an \emph{initial value problem}.  Let us expand on this point.  Hamilton's Principle in the above derivation asks: if we know the initial position at the initial time of our point particle and \emph{a priori} know the final position of the particle at the final time of interest, what is the path the particle takes?  In contrast, in a realistic experiment, generally speaking it is only the \emph{initial conditions} of the initial position and initial velocity at the initial time that are known.  To know the final location of the particle at some future time $t_f$ is to possess \emph{acausal} information, a fundamental violation of Einstein's Special Relativity \cite{Einstein1905}.  Further, unlike initial value problems (which are well-posed, in the sense that the Picard-Lindel\"of Theorem \cite{CoddingtonLevinson1955} guarantees the existence and uniqueness of the solution), boundary value problems can have one, zero, or many solutions \cite{CoddingtonLevinson1955}, contrary to our expectation and experience that experimental measurements can be made and are reproducible \cite{Popper1959}.  

The second conceptual issue is as follows.  We know that the velocity of the particle is related to the position of the particle through time differentiation,
\begin{align}
    \dot q^i(t) = \frac{d}{dt}\big[ q^i(t) \big].
\end{align}
Why, then, do we treat the variations in the velocity of the particle $\delta\dot q^i(t)$ as independent of the variations of the path of the particle $\delta q^i(t)$ in the derivation of the Euler-Lagrange Equations?  In a similar way, in the derivation of Hamilton's Equations from Hamilton's Principle, one treats the variations in the momentum $\delta p_i(t)$ as independent of the variations in the paths $\delta q^i(t)$, even though the defintion of momentum
\begin{align}
    p_i \equiv \frac{\partial L}{\partial \dot q^i}
\end{align}
implies that the momentum is a function of the path and the velocity, $p_i\big(q^i(t),\dot q^i(t)\big)$.

Related to this second conceptual issue is the confusion and open questions related to the treatment of constraints in Hamilton's formulation.  Holonomic constraints $f_k\big( q^i(t),t \big) = 0$, such as constraining the particle to move on a circle $x^2(t)+y^2(t)-R^2=0$, depend only on the coordinates $q^i(t)$.  Hamilton's Principle in this case is modified such that the variations in the path must satisfy the constraint equations, $f_k\big( q^i(t) + \delta q^i(t),t \big) = 0$.  The variations are not all independent and arbitrary, but one may still find solutions to the problem.  In particular, the transposition rule \cref{eq:transpositionrule} remains unchanged.  However, for non-holonomic constraints $g_k\big( q^i(t),\dot q^i(t),t \big)=0$, such as for the problem of rolling without slipping \cite{Flannery2005_enigma_nonholonomic,Flannery2011DAlembertLagrange}, a general formulation of Hamilton's Principle eluded physicists until only recently \cite{RothkopfHorowitz2024APS}. Further, for non-holonomic constraints, the transposition rule \cref{eq:transpositionrule} appears to be modified \cite{Flannery2011DAlembertLagrange}, which therefore modifies the symplectic structure of phase space \cite{de_Le_n_2024,Horowitz:2024eea}.  Modifying this symplectic structure of phase space has unknown consequences for the quantum mechanics of systems with non-holonomic constraints, such as quantum nano-carts \cite{FernandezRadhakrishnan2018} and gauge theories under the 't Hooft-Veltman gauge \cite{Bert:2025ffu}.

In addition to the above conceptual issues related to the derivation of the Euler-Lagrange Equations from Hamilton's Principle is the limitation that Hamilton's formulation of classical mechanics requires that forces be derived from potentials, which is to say that the system must be conservative; i.e.\ dissipative systems are outside the scope of the treatment.  In a recent work \cite{Galley:2012hx}, Galley proposed a new formulation of classical mechanics.  In Galley's Principle, one also considers the variation of an action, but with two crucial changes: first, the number of degrees of freedom for the system are doubled, and, second, a new term is added to the action, which can incorporate the effects of dissipation.  In particular, Galley proposed that one should consider an action dependent on two paths, $\vec q_1(t)$ and $\vec q_2(t)$, and on a term $K\big( \vec q_a(t),\dot{\vec q}_a(t),t \big)$ that depends on both paths $\{\vec q_a\}^2_{a=1}$,
\begin{align}
    S_G \equiv \int dt L\big( \vec q_1(t), \dot{\vec q}_1(t) \big) - L\big( \vec q_2(t), \dot{\vec q}_2(t) \big) + K\big( \vec q_a(t),\dot{\vec q}_a(t),t \big).
\end{align}
Utilizing the change of coordinates
\begin{align}
    \begin{aligned}
        \vec q_+(t) & \equiv \frac12\big(\vec q_1(t) + \vec q_2(t)\big) \\
        \vec q_-(t) & \equiv \vec q_1(t) - \vec q_2(t),
    \end{aligned}
\end{align}
Galley then proposed that the classical path is one for which the modified action is extremized
\begin{align}
    \delta S_G|_{\substack{\vec q_{+,cl}(t) \\ \vec q_{-,cl}(t)}} = 0,
\end{align}
where the variations in the paths are restricted such that
\begin{align}
    \delta x_-(t_f) & = 0 \nonumber\\
    \delta x_-(t_i) & = 0 \\
    \delta x_+(t_i) & = 0. \nonumber
\end{align}

In principle, Galley's formulation appeared to simultaneously resolve two of the three above issues: the formulation appeared to allow a variational principle of classical mechanics as an initial value problem and that could incorporate dissipative effects.

However, an attempt to implement Galley's Principle in a numerical simulation of various point particle mechanics problems \cite{Rothkopf:2023ljz} had two puzzling features.  First, the trajectory found numerically that extremized the action had unphysical jumps at the final time slice.  Second, the minus coordinate trajectory always solved to exactly zero.

At the same time, Galley's doubling of the degrees of freedom was extremely reminiscent of the doubling of the degrees of freedom in the Schwinger-Keldysh/closed-time path/in-in formulation of the time dependence of the expectation value of an operator in quantum mechanics \cite{Schwinger1961,Keldysh1965}.  The Schwinger-Keldysh, or closed time path, formulation of quantum mechanics is used in a wide range of applications, including non-equilibrium statistical mechanics and quantum field theory \cite{Berges:2004yj,Rammer2007,BalzerBonitz2013}, condensed matter systems \cite{Kamenev2011}, quantum field theory at finite temperature \cite{Landsman:1986uw,Das:1997gg,Laine:2016hma}, cosmology \cite{Jordan1986,Weinberg2005}, and in strongly coupled systems and holography \cite{Casalderrey-Solana:2011dxg}, to name a few.  In the context of these fields of study, it is generally said that the minus degrees of freedom are the quantum fluctuations about the classical path of the system, and that the classical $\hbar\rightarrow0$ limit is found by setting the minus degrees of freedom to zero (by hand).

Galley's work also strongly resembled that of the open quantum system community.  While Feynman and Vernon \cite{Feynman:1963fq} and Caldeira and Leggett \cite{Caldeira:1982iu} did not quite consider an in-in expectation value, there are clear connections.  For example, Caldeira and Leggett specifically discuss their work in the context of dissipative systems, interpreting the effective action of Morse and Feshbach \cite{MorseFeshbach1953} as having a $K$ term, in the conventions of Galley, as a function of \emph{both} forward and backwards paths in the Schwinger-Keldysh formalism.  However, it does not appear that Caldeira-Leggett explicitly connected their work to a classical variational action principle.  

In what follows, we will accomplish the following.  1) We will provide a rigorous derivation of the $\hbar\rightarrow0$ limit of the Schwinger-Keldysh formulation of non-relativistic point particle quantum mechanics in order to derive a variational action principle for the initial value problem of classical mechanics for Hamilton's Equations and for the Euler-Lagrange Equations.  2) In so carefully taking this limit, we will find that Galley's Principle should be modified as follows.  We show explicitly that for non-dissipative systems the action to be varied should rather be Hamilton's Revised Action,
\begin{align}
    S_{HR} \equiv \vec p_0 \cdot \vec q_-(t) + \int dt L\big( \vec q_1(t), \dot{\vec q}_1(t) \big) - L\big( \vec q_2(t), \dot{\vec q}_2(t) \big),
\end{align}
where the classical path is given by
\begin{align}
    \delta S_{HR}|_{\substack{\vec q_{+,cl}(t) \\ \vec q_{-,cl}(t)}} = 0,
\end{align}
where the variations in the paths are restricted such that
\begin{align}
    \begin{aligned}
        \delta x_-(t_f) & = 0 \\
        \delta x_+(t_i) & = 0
    \end{aligned}
\end{align}
and $\vec p_0$ is the initial momentum of the particle.  While we have not yet explicitly checked for dissipative systems, presumably the above is simply altered by including the dissipative term $K\big(\vec q_a(t),\dot{\vec q}_a(t)\big)$.  3) We will then show how to correctly vary a Lagrangian that depends on both the position $\vec q(t)$ and the velocity of the particle $\dot{\vec q}(t)$.  We will thus fully resolve the two conceptual issues of classical point particle mechanics noted above.  4) In the process of deriving the above results, we will also see how, when taking the $\hbar\rightarrow0$ limit of the Schwinger-Keldysh path integral, the minus degrees of freedom are all set to zero by the equations of motion induced by the $\hbar\rightarrow0$ limit; i.e.\ we see that one does not have to take the minus degrees of freedom to zero by hand when taking the classical limit.  5) We will first compute the above in a single Cartesian coordinate.  When generalizing the above results to $d$ (not necessarily Cartesian) dimensions, we will rigorously derive the expression for the Trotterization of a quantum term that depends on both position and momentum that does not rely on \emph{ad hoc} prescriptions or introducing new quantum mechanics principles.  We believe that the techniques developed in considering the Schwinger-Keldysh path integral over a particle's position and velocity will in future work help resolve the open questions related to the classical and quantum mechanics of general non-holonomic systems and also allow for computing the $\hbar\rightarrow0$ limit of dissipative systems.  

\section{Notation and Conventions}
\label{sec:notation}

The bulk of our work will be in one Cartesian dimension, $x$.  We will describe the position of our single particle as a function of time as $x(t)$.  We will take our Lagrangian to be of the form
\begin{align}
    L = \frac12m\left( \frac{dx}{dt}\right)^2 - V(x).
\end{align}
The associated momentum of this particle will be $p(t)$, where the momentum of the particle as a function of time, $p(t)$, is proportional to the mass of the particle and the time derivative of the position of the particle evaluated at time $t$, $p(t) = \left.\frac{\partial L}{\partial[\frac{dx}{dt}]}\right|_t = m \left.\frac{dx}{dt}\right|_t$.  

In order to facilitate and clarify the discussion relating to the variation of the action with respect to the path $x(t)$ and to the velocity $\dot x(t)$, we will carefully distinguish between the time derivative of the path evaluated at time $t$,
\begin{align}
    \left.\frac{dx}{dt}\right|_t = \frac{d}{dt}\big[ x(t) \big] = \left. \frac{d}{dt}\big[ x \big]\right|_t
\end{align}
and the symbol $\dot x(t)$.  We will ultimately show that, in the $\hbar\rightarrow0$ limit  of the path integrals we consider,
\begin{align}
    \dot x(t) = \left.\frac{dx}{dt}\right|_t,
\end{align}
which perhaps at this stage seems unnecessarily pedantic, but we will see that in so doing makes the classical variational procedure unambiguous. 

There are many formulations of the variation of a path in classical mechanics \cite{goldstein2002classical,Flannery2011DAlembertLagrange}.  For example, \cite{goldstein2002classical,Galley:2012hx} uses a common formulation in which paths are considered as a function not only of the time $t$ but also of a small parameter $\alpha>0$,
\begin{align}
    x(t;\alpha) = x(t;0) + \alpha \eta(t). \nonumber
\end{align}
$\eta(t)$ is some real function, and the classical motion is found by expanding all quantities order-by-order in $\alpha$ and ensuring that the linear in $\alpha$ terms vanish.  

In order to reduce notational clutter, we will follow \cite{Flannery2011DAlembertLagrange}, in which the variation of the path $x(t)$ is denoted by $\delta x(t)$,
\begin{align}
    x(t) = x_{cl}(t) + \delta x(t). 
\end{align}
In the above, we consider $\delta x$ as a single function name; i.e.\ one considers $(\delta x)(t)$.  In a way that should be clear from context, we will also refer to the variation of an action $S$ as $\delta S$.  In the case of the variation of the action, $\delta$ is acting as an operation; e.g.,
\begin{align}
    \delta S\Big[ x(t), \frac{d}{dt}\big[ x(t) \big] \Big] = S\Big[x(t) + \delta x(t), \frac{d}{dt}\big[ x(t) + \delta x(t) \big]\Big] - S\Big[ x(t), \frac{d}{dt}\big[ x(t) \big] \Big].
\end{align}

We note that we will determine all final results in terms of the final time $t_f$, as opposed to the more usual solutions as a function of time $t$.  We made this choice to avoid the notational clutter of having integrals over $t'$ in our actions.

In order to fix terminology, we note that will consider three types of path integrals, first over discrete time slices and then in the continuum limit: a phase space path integral over $x_\pm(t)$ and $p_\pm(t)$; a position space path integral over $x_\pm(t)$; and a configuration space path integral over $x_\pm(t)$, $\dot x_\pm(t)$, and over Lagrange multiplier functions $\lambda_\pm(t)$.  

\section{Hamilton's Equations from Schwinger-Keldysh}
\label{sec:hamiltonsequations}
\subsection{Discrete Case}
\label{sec:discreteham}
Let us consider the physical problem of the time evolution of the expectation value of the position of a quantum wavepacket in a non-relativistic system under the influence of forces derivable from a potential depending on position only.

We start with the definition of the expectation value of the position operator at time $t_f$,
\begin{align}
    \label{eq:expectationvaluedefinition}
    \langle \hat x\rangle(t_f) = \langle\psi(t_i)|\hat U^\dagger_{\hat H}(t_f,t_i)\,\hat x \,\hat U_{\hat H}(t_f,t_i)|\psi(t_0)\rangle,
\end{align}
where $\hat U_{\hat H}(t_f,t_i)$ is the usual time evolution operator,
\begin{align}
    \hat U_{\hat H}(t_f,t_i) & = e^{{-}i\hat H(t_f-t_i)}.
\end{align}
We will take for simplicity
\begin{align}
    \hat H(\hat x,\hat p) & = \frac{\hat p^2}{2m} + V(\hat x).
\end{align}
We will show how to consider a much more general Hamiltonian associated with generalized coordinates in App.\ \ref{sec:generalizedcoordinates}.  We will further take our initial state to be a pure Gaussian wavepacket,
\begin{align}
    \label{eq:initialket}
    |\psi(t_i)\rangle & = \int_{-\infinity}^\infinity dx\left( \frac{1}{2\pi\sigma_x^2} \right)^{1/4}e^{\frac{ip_0x}{\hbar}}e^{-\frac{(x-x_0)^2}{4\sigma_x^2}}|x\rangle.
\end{align}
One could take the initial state to be more general, e.g.\ a density matrix.  We choose to consider ``just'' the Gaussian wavepacket for simplicity.  We will find that our results look exceptionally natural in this case.  Presumably any initial state that includes smearing in position and momentum and that satisfies a minimal uncertainty relation will yield similar results, although we have not performed any explicit checks.

One may readily work out for our Gaussian initial wavepacket that
\begin{align}
    \begin{aligned}
        \langle\psi(t_i)|\psi(t_i)\rangle & = 1 \\
        \langle\hat x\rangle(t_i) & = x_0 \\
        \langle(\hat x-x_0)^2\rangle(t_i) & = \sigma_x^2 \\
        \langle\hat p\rangle(t_i) & = p_0 \\
        \label{eq:momentumvariancesquared}
        \langle(\hat p-p_0)^2\rangle(t_i) & = \frac{\hbar^2}{4\sigma_x^2}.
    \end{aligned}
\end{align}
Crucially, \cref{eq:momentumvariancesquared} implies that the width in the momentum $\sigma_p$ is proportional to $\hbar$ and inversely related to the width in the position of the wavepacket:
\begin{align}
    \label{eq:momentumspreadhbar}
    \sigma_p = \frac{\hbar}{2\sigma_x}.
\end{align}
\Cref{eq:momentumspreadhbar} tells us that we need to carefully order our limits as we approach classical mechanics: we must first take $\hbar\rightarrow0$ and \emph{only then} take $\sigma_x\rightarrow0$.

Let us now consider the time evolution of the expectation value of the position, \cref{eq:expectationvaluedefinition}.  For notational simplicity, we will suppress $\hbar$ in the intermediate steps of the calculation and restore $\hbar$ at the end:
\begin{multline}
    \label{eq:expectationvalue120}
    \langle x\rangle (t_f)
    = \int dx_{1,0}\left( \frac{1}{2\pi\sigma_x^2} \right)^{1/4}e^{ip_0x_{1,0}}e^{-\frac{(x_{1,0}-x_0)^2}{4\sigma_x^2}} \\ \int dx_{2,0}\left( \frac{1}{2\pi\sigma_x^2} \right)^{1/4}e^{-ip_0x_{2,0}}e^{-\frac{(x_{2,0}-x_0)^2}{4\sigma_x^2}} 
    \langle x_{2,0}|e^{i\left( \frac{\hat p^2}{2m} + V(\hat x) \right)\Delta t}\,\hat x\,e^{-i\left( \frac{\hat p^2}{2m} + V(\hat x) \right)\Delta t}|x_{1,0}\rangle,
\end{multline}
where $\Delta t\equiv t_f-t_i$.  Define $\epsilon\equiv\Delta t/N$, and let us Trotterize \cite{Trotter:1959ytf}: 
\begin{align}
    \label{eq:expectationvalue121}
    \langle x\rangle (t_f)
    & = \lim_{N\rightarrow\infinity}\int dx_{1,0}\left( \frac{1}{2\pi\sigma_x^2} \right)^{1/4}e^{ip_0x_{1,0}}e^{-\frac{(x_{1,0}-x_0)^2}{4\sigma_x^2}}\int dx_{2,0}\left( \frac{1}{2\pi\sigma_x^2} \right)^{1/4}e^{-ip_0x_{2,0}}e^{-\frac{(x_{2,0}-x_0)^2}{4\sigma_x^2}} \nonumber\\
    & \quad \langle x_{2,0}|e^{iV(\hat x)\epsilon}\int dp_{2,\frac12}|p_{2,\frac12}\rangle\langle p_{2,\frac12}|e^{i\frac{\hat p^2}{2m}\epsilon}\int dx_{2,1}|x_{2,1}\rangle\langle x_{2,1}|\cdots\int dx_{2,N}|x_{2,N}\rangle\langle x_{2,N}|\hat x \nonumber\\
    & \quad \int dx_{1,N}|x_{1,N}\rangle\langle x_{1,N}|\cdots\int dx_{1,1}|x_{1,1}\rangle\langle x_{1,1}|e^{-i\frac{\hat p^2}{2m}\epsilon}\int dp_{1,\frac12}|p_{1,\frac12}\rangle\langle p_{1,\frac12}|e^{-iV(\hat x)\epsilon}|x_{1,0}\rangle \nonumber\\
    & \quad + \mathcal O(\epsilon^2).
\end{align}
Notice that we place the momentum eigenstate insertions at the half time steps between the position eigenstate insertions.  In principle, the exact location in time of the momentum eigenstate insertions can be anytime between the adjacent position eigenstate times, inclusive.  We make this specific choice of an exactly half time step for the location of the momentum time steps because numerical evaluation of the equations of motion we will derive show that this choice gives a much more accurate comparison to the continuous time solution.  In fact, this choice gives the exactly correct discrete time step placement for the momentum when compared to the continuous time solution when the original Trotterization is performed to $\mathcal O(\epsilon^3)$ through Strang splitting \cite{Strang1968} and for potentials at most linear in position.  

For canonically conjugate pairs of variables, one has from the Dirac Quantization Condition
\begin{align}
    [\hat x,\hat p] = i,
\end{align}
that
\begin{align}
    \langle x|p\rangle = \frac{1}{\sqrt{2\pi}}e^{ipx}.
\end{align}
We then have that \cref{eq:expectationvalue121} is
\begin{align}
    \langle \hat x\rangle (t_f)
    \label{eq:expectationvalue122}
    & = \lim_{N\rightarrow\infinity}\prod_{n_x=0}^{N}\int dx_{1,n_x}dx_{2,n_x}\prod_{n_p=0}^{N-1}\int \frac{dp_{1,n_p+\frac12}}{2\pi}\frac{dp_{2,n_p+\frac12}}{2\pi} \nonumber\\
    & \quad \left( \frac{1}{2\pi\sigma_x^2} \right)^{1/4}e^{{-}\frac{(x_{2,0}-x_0)^2}{4\sigma_x^2}}e^{-ip_0x_{2,0}}e^{{-}i\epsilon\sum_{k_2=0}^{N-1}p_{2,k_2+\frac12}\frac{x_{2,k_2+1}-x_{2,k_2}}{\epsilon} - H(p_{2,k_2+\frac12},x_{2,k_2}) } \nonumber\\
    & \quad \left( \frac{1}{2\pi\sigma_x^2} \right)^{1/4}e^{{-}\frac{(x_{1,0}-x_0)^2}{4\sigma_x^2}}e^{ip_0x_{1,0}}e^{i\epsilon\sum_{k_1=0}^{N-1}p_{1,k_1+\frac12}\frac{x_{1,k_1+1}-x_{1,k_1}}{\epsilon} - H(p_{1,k_1+\frac12},x_{1,k_1}) } \nonumber\\
    & \quad x_{1,N}\delta(x_{1,N}-x_{2,N}) + \mathcal O(\epsilon^2), 
\end{align}
where
\begin{align}
    H(p_{i,k_i+\frac12},x_{i,k_i}) & \equiv \frac{p_{i,k_i+\frac12}^2}{2m} + V(x_{i,k_i}).
\end{align}
Notice in \cref{eq:expectationvalue122} the natural emergence of a \emph{connecting condition} at the final time slice $N$ between the forward ($x_1$) and backward ($x_2$) time evolution branches of the Schwinger-Keldysh contour, setting $x_{1,N} = x_{2,N}$ from the quantum matrix element $\langle x_{2,N}|\hat x|x_{1,N}\rangle = x_{1,N}\delta(x_{1,N}-x_{2,N})$.  Notice further that there is \emph{no} connecting condition between the momenta.  Notice, finally, there is an \emph{unnatural} difference in the number of integrals over position and the number of integrals over momentum; we will see how to resolve this unnaturalness as we proceed.

We would now like to change variables from the $1/2$ coordinates to $+/-$ coordinates, defined as
\begin{align}
    \label{eq:plusminuscoordsdefinition}
    \begin{aligned}
    x_{+,k}(t) & \equiv \frac12\big( x_{1,k}(t) + x_{2,k}(t) \big); & p_{+,k+\frac12} & \equiv \frac12\big( p_{1,k+\frac12}(t) + p_{2,k+\frac12}(t) \big) \\[5pt]
    x_{-,k}(t) & \equiv x_{1,k}(t) - x_{2,k}(t); & p_{-,k+\frac12} & \equiv p_{1,k+\frac12}(t) - p_{2,k+\frac12}(t).
    \end{aligned}
\end{align}
This change of coordinates is especially useful and, as we will see, the $+/-$ coordinates are the natural ones to use as we seek the classical limit.  It is often written that the minus coordinates should be thought of as the quantum fluctuations around the classical path that the plus coordinates follow.  What we will see is that this interpretation is \emph{not} correct, as both the plus and minus coordinates contain \emph{both} a classical trajectory and quantum fluctuations around that classical trajectory.  We will show that the correct understanding, rather, is that the classical limit forces the minus coordinates to be identically zero for all time, $x_-(t)=p_-(t)=0$, while the plus coordinates follow Hamilton's Equations of Motion as an initial value problem.  

Note that one should not confuse the above definition of the $+/-$ coordinates \cref{eq:plusminuscoordsdefinition}, which mix the forward and backward labels of the \emph{eigenstates} associated with a single canonically conjugate position and momentum operator pair $\hat x$ and $\hat p$, with the $+/-$ lightcone coordinates often used in high-energy physics, which mix the $t$ and $z$ coordinates of Minkowski four-vectors.  In particular, it does not make sense to consider in our formalism $[\hat x_i,\hat p_j]$, $[\hat x_\pm,\hat p_\pm]$, or $[\hat x_\pm,\hat p_\mp]$, because there is---in fact---only one $\hat x$ and one $\hat p$ for our single non-relativistic particle moving in one dimension.

The Jacobian associated with the change of variables \cref{eq:plusminuscoordsdefinition} is 1; we thus have that \cref{eq:expectationvalue121} becomes
\begin{align}
    \label{eq:expectationvaluepm0}
    \langle \hat x\rangle (t_f)
    & = \lim_{N\rightarrow\infinity} \int dx_{+,0}dx_{-,N}\prod_{n=0}^{N-1}dx_{+,n+1}dx_{-,n}\frac{dp_{+,n+\frac12}}{2\pi\hbar}\frac{dp_{-,n+\frac12}}{2\pi\hbar}\nonumber\\
    & \quad \left( \frac{1}{2\pi\sigma_x^2} \right)^{1/2}e^{{-}\frac{\left(x_{+ ,0} - x_0\right)^2}{2\sigma_x^2}}e^{{-}\frac{x_{-,0}^2}{8\sigma_x^2}}e^{\frac{i}{\hbar}p_0x_{-,0}} \nonumber\\
    & \quad \exp\left\{\frac{i\epsilon}{\hbar}\sum_{k=0}^{N-1}\left[ p_{+,k+\frac12}\frac{x_{-,k+1}-x_{-,k}}{\epsilon} + p_{-,k+\frac12}\frac{x_{+,k+1}-x_{+,k}}{\epsilon} \right.\right. \nonumber\\
    & \quad \left.\left.+ H\left(p_{+ ,k+\frac12} - \frac12p_{-,k+\frac12},x_{+ ,k} - \frac12x_{-,k}\right)  - H\left(p_{+ ,k+\frac12} + \frac12p_{-,k+\frac12},x_{+ ,k} + \frac12x_{-,k}\right)\right] \right\} \nonumber\\
    & \quad x_{+ ,N}\delta(x_{-,N}) + \mathcal{O}(\epsilon^2).
\end{align}
One can see how the connecting condition between the positions of the forward and backward $1/2$ paths has been converted to the condition that $x_{-,N}=0$.  Notice importantly how the integrals over positions $dx_{\pm,n}$ naturally pair up with the integrals over the momenta $dp_{\pm,n+\frac12}$ \emph{except} for the two leftover integrals over positions $dx_{+,0}$ and $dx_{-,N}$.  In our next step, we will understand how to naturally pair up these leftover integrals of positions with integrals over two new momenta, in the form of Lagrange multipliers.  

In \cref{eq:expectationvaluepm0}, we have restored the $\hbar$'s to better motivate our next steps.  In particular, we plan to utilize the Method of Stationary Phase, which will be valid in the $\hbar\rightarrow0$ limit.  The Method of Stationary Phase for a multidimensional integral is given in general by \cite{Hormander2003}
\begin{multline}
    \label{eq:stationaryphase}
    \int_{\mathbb R^n} d^n x\,g(x)e^{i\,\phi(x)/a}  
    = \\ \sum_{x_0} \big|\det \big(\Hess(\phi)|_{x_0}\big)\big|^{-1/2}e^{i\frac{\pi}{4}\sgn\big(\Hess(\phi)|_{x_0}\big)}(2\pi \, a)^{n/2}g(x_0)e^{i \phi(x_0)/a} + o(a^{n/2}),
\end{multline}
where the sum is over all critical points $x_0$ such that $\nabla_x \phi|_{x_0}=0$ and $\Hess(\phi)|_{x_0}$ is the Hessian matrix of second derivatives of $\phi$ evaluated at the critical point $x_0$.  Here $\sgn$ is the signature of a matrix, which is the difference in the number of positive eigenvalues and the number of negative eigenvalues of the matrix. We want to emphasize that we will not simply be considering the path that extremizes the action; rather, we are considering rigorously the $\hbar\rightarrow0$ limit of the full path integral utilizing the Method of Stationary Phase.  The Method of Stationary Phase, which is rigorously providing for us the leading order in $\hbar$ result of the path integrals, demands that we consider the paths that leave the action stationary.

We would thus like to utilize the following identities to replace the delta function enforcing the connecting condition at the final time slice and the Gaussian enforcing the initial wavepacket condition at the initial time slice with objects that naturally enter the phase of the path integral:
\begin{align}
    \begin{aligned}
        e^{-\alpha x^2} & = \frac{1}{\sqrt{4\pi\alpha}}\int_{-\infinity}^\infinity d\lambda e^{-\frac{\lambda^2}{4\alpha}}e^{+i\lambda x}, \qquad \alpha>0 \\[10pt]
        \delta(x) & = \frac{1}{2\pi}\int_{-\infinity}^\infinity d\lambda e^{-i\lambda x}.
    \end{aligned}
\end{align}
Once inserted into \cref{eq:expectationvaluepm0}, we should think of the $\lambda$'s as 1) Lagrange multipliers and 2) as new momenta that will pair up with our leftover position integrals, a statement that we will make precise in the next set of equations.  Recall again that the positions and momenta here are \emph{not} canonically conjugate pairs; rather, there is only one position operator $\hat x$ and one momentum operator $\hat p$, and this pairing up is between the eigenvalues of the eigenstates of this one pair of canonically conjugate operators.  

We thus find that \cref{eq:expectationvaluepm0} becomes
\begin{align}
    \label{eq:expectationvaluepm1}
    \langle \hat x\rangle (t_f)
    & = \lim_{N\rightarrow\infinity} \int dx_{+,0}\frac{d\lambda_{-,0}}{2\pi\hbar}dx_{-,N}\frac{d\lambda_{+,N}}{2\pi\hbar}\prod_{n=0}^{N-1}dx_{+,n+1}dx_{-,n}\frac{dp_{+,n+\frac12}}{2\pi\hbar}\frac{dp_{-,n+\frac12}}{2\pi\hbar}\nonumber\\
    & \quad  e^{-\frac{x_{-,0}^2}{8\sigma_x^2}}e^{{-}\frac{\sigma_x^2\lambda_{-,0}^2}{2\hbar^2}}x_{+ ,N}e^{\frac i\hbar S_{HR}[p_{\pm,k+\frac12},x_{\pm,k},\lambda_{\pm}]}+\mathcal{O}(\epsilon^2), \\
    \label{eq:discreteHRaction}
    S_{HR}[p_{\pm,k+\frac12},x_{\pm,k},\lambda_\pm]
    & \equiv \exp\left\{\frac{i}{\hbar}\left[\lambda_{-,0}(x_{+,0}-x_0)-\lambda_{+,N}\,x_{-,N}+p_0\,x_{-,0} \phantom{V\left(x_{+ ,k} + \frac12x_{-,k}\right)} \right.\right. \nonumber\\
    & \qquad\qquad + \epsilon\sum_{k=0}^{N-1}\left( p_{+,k+\frac12}\frac{x_{-,k+1}-x_{-,k}}{\epsilon} + p_{-,k+\frac12}\frac{x_{+,k+1}-x_{+,k}}{\epsilon}\right.\nonumber\\
    & \qquad\qquad\qquad + H\Bigl(p_{+ ,k+\frac12} - \frac12p_{-,k+\frac12},x_{+ ,k} - \frac12x_{-,k}\Bigr) \nonumber\\
    & \qquad\qquad\qquad - \left.\left.\left.H\Bigl(p_{+ ,k+\frac12} + \frac12p_{-,k+\frac12},x_{+ ,k} + \frac12x_{-,k}\Bigr)\right) \right] \right\}.
\end{align}

We see in \cref{eq:expectationvaluepm1} that the $\lambda_\pm$ act as Lagrange multipliers that correct the unnatural imbalance in the number $N$ of $x_\pm$ integrals and the number $N-1$ of $p_{\pm}$ integrals.  (The importance of balancing out the number of these integrals is seen most acutely in the $\hbar$ counting, which is very natural with the addition of the $\lambda_\pm$ integrals, where the $\lambda_\pm$ act as momenta.)  It is easy to see that the delta function $\delta(x_{-,N})$ completely restricts the $x_{-,N}$ degree of freedom; we could have collapsed that delta function with the $\int dx_{-,N}$.  One can further intuitively see that for $\sigma_x\rightarrow0$ the Gaussian $\exp[-(x_{+,0}^2-x_0)^2/2\sigma_x^2]$ looks like a Dirac delta function centered at $x_{+,0}=x_0$ that could be used to collapse the $\int dx_{+,0}$ integral.  However, in a way that is reminiscent of the Kugo-Ojima mechanism \cite{Kugo:1979gm} we will find it much more natural and useful to increase the number of degrees of freedom by introducing the Lagrange multipliers $\lambda_\pm$ as new, independent degrees of freedom that enforce the conditions on $x_{+,0}$ and $x_{-,N}$ and keeping the integrals over $x_{+,0}$ and $x_{-,N}$ rather than decreasing the number of degrees of freedom by collapsing the $x_{+,0},\,x_{-,N}$ integrals.  From the structure of the phase in \cref{eq:expectationvaluepm1}, it is clear that $\lambda_{-,0}$ pairs with $x_{+,0}$ while $\lambda_{+,N}$ pairs with $x_{-,N}$.  (We will see in a moment that $\lambda_{+,N}$ is the final classical momentum of the particle at the final time slice $N$ while $\lambda_{-,0}$ is the ``final'' minus momentum from what we will see is the backwards time propagation of the minus coordinates.)

For completeness, let us note a subtlety.  If one carefully compares our phase space path integral with Lagrange multipliers \cref{eq:expectationvaluepm1} to the general Method of Stationary Phase formula \cref{eq:stationaryphase}, one may notice an inconsistency.  In \cref{eq:stationaryphase}, the non-phase part of the integrand $g(x)$ is independent of the small parameter $a$.  However, in our phase space path integral \cref{eq:expectationvaluepm1}, the non-phase part of the integrand explicitly depends on $\hbar$ (in fact, $\hbar^2$).  One might then worry that the general formula \cref{eq:stationaryphase} does not apply or may need to be modified in our case.  We will delay a careful consideration of this subtlety until \cref{sec:stationaryphase}, but note that the subtlety will not change our approach or final result.

Defining the phase in \cref{eq:expectationvaluepm1} as $S_{HR}$ (for the Schwinger-Keldysh action), the Method of Stationary Phase implies that in the classical $\hbar\rightarrow0$ limit, the leading order contribution from the integrals that yield the expectation value of $\hat x(t)$ will come from those values of $x_{\pm,k},\,p_{\pm,k+\frac12},$ and $\lambda_\pm$ that extremize $S_{HR}$.  I.e.\ we seek to find those values of $x_{\pm,k},\,p_{\pm,k+\frac12},$ and $\lambda_\pm$ such that $\delta S_{HR} = 0$.  Varying $S_{HR}$ with respect to $x_{\pm,k},\,p_{\pm,k+\frac12},$ and $\lambda_\pm$, we find
\begin{align}
    \label{eq:discreteHRpmvariation0}
    \delta S_{HR}[p_{\pm,k+\frac12},x_{\pm,k},\lambda_\pm]
    & = \delta \lambda_{-,0}(x_{+,0}-x_0) + \lambda_{-,0}\delta x_{+,0} - \delta \lambda_{+,N}x_{-,N} - \lambda_{+,N}\delta x_{-,N} + p_0\delta x_{-,0} \nonumber\\
    & \quad +\epsilon\sum_{k=0}^{N-1}\left[ \delta p_{+,k+\frac12}\frac{x_{-,k+1}-x_{-,k}}{\epsilon} + p_{+,k+\frac12}\frac{\delta x_{-,k+1}-\delta x_{-,k}}{\epsilon} \right. \nonumber\\
    & \quad + \delta p_{-,k+\frac12} \frac{x_{+,k+1}-x_{+,k}}{\epsilon} + p_{-,k+\frac12}\frac{\delta x_{+,k+1} - \delta x_{+,k}}{\epsilon} \nonumber\\
    & \quad + \left.\frac{\partial H}{\partial p}\right|_{-,k}\left( \delta p_{+,k+\frac12} - \frac12\delta p_{-,k+\frac12} \right) + \left.\frac{\partial H}{\partial x}\right|_{+,k}\left( \delta x_{+,k} - \frac12\delta x_{-,k} \right) \nonumber\\
    & \quad \left. - \left.\frac{\partial H}{\partial p}\right|_{+,k}\left( \delta p_{+,k+\frac12} + \frac12\delta p_{-,k+\frac12} \right) - \left.\frac{\partial H}{\partial x}\right|_{+,k}\left( \delta x_{+,k} + \frac12\delta x_{-,k} \right) \right],
\end{align}
where
\begin{align}
    \begin{aligned}
        \left.\frac{\partial H}{\partial p}\right|_{\pm,k} & \equiv \left.\frac{\partial H}{\partial p}\right|_{p = p_{+ ,k+\frac12} \pm \frac12p_{-,k+\frac12},\,x = x_{+ ,k} \pm \frac12x_{-,k}},
    \end{aligned}
\end{align}
and similar for $\partial H/\partial x|_{\pm,k}$.

To proceed, we perform some index manipulation:
\begin{align}
    \label{eq:indexmanipulation}
    \sum_{k=0}^{N-1} p_{+,k+\frac12}(\delta x_{-,k+1} - \delta x_{-,k}) 
    & = \sum_{k=1}^{N} p_{+,k-1+\frac12}\delta x_{-,k} - \sum_{k=0}^{N-1} p_{+,k+\frac12}\delta x_{-,k} \nonumber\\
    & = p_{+,N-1+\frac12}\delta x_{-,N} - p_{+,\frac12}\delta x_{-,0} \nonumber\\
    & \pushright{+ \sum_{k=1}^{N-1} (p_{+,k-1+\frac12} - p_{+,k+\frac12})\delta x_{-,k}.}
\end{align}
In the second line above, \cref{eq:indexmanipulation}, we intentionally kept the awkward-looking $k-1+\frac12$ index for the momentum $p$.  We kept this awkward-looking index instead of using $k-\frac12$ because in the case when there are multiple eigenvalues within one time step, e.g.\ in the case of considering a second order splitting or, as noted below, for the case of non-Cartesian and/or non-Euclidean coordinates, each variable associated with each shifted index will evolve according to its own equation of motion in the $\hbar\rightarrow0$ limit.  See, e.g., App.\ \ref{sec:generalizedcoordinates}, where the $p_{k+\frac14}$ evolves differently from the $p_{k+\frac34}$, and it would not make sense to take $p_{k-1+\frac14} = p_{k-\frac34}$.   

Therefore,
\begin{align}
    \label{eq:discreteHRpmvariation1}
    \delta S_{HR}[p_{\pm,k+\frac12},x_{\pm,k},\lambda_\pm]
    & = \delta \lambda_{-,0}[x_{+,0}-x_0] \nonumber\\
    & \quad + \delta x_{-,0}\left[ p_0 - p_{+,\frac12}-\frac\epsilon2\left(\left.\frac{\partial H}{\partial x}\right|_{+,0} + \left.\frac{\partial H}{\partial x}\right|_{-,0} \right) \right] \nonumber\\
    & \quad + \sum_{k=0}^{N-1}\delta p_{-,k+\frac12}\left[ x_{+,k+1}-x_{+,k} - \frac\epsilon2\left( \left.\frac{\partial H}{\partial p}\right|_{+,k} + \left.\frac{\partial H}{\partial p}\right|_{-,k} \right) \right] \nonumber\\
    & \quad + \sum_{k=1}^{N-1}\delta x_{-,k}\left[ p_{+,k-1+\frac12}-p_{+,k+\frac12} - \frac\epsilon2\left(\left.\frac{\partial H}{\partial x}\right|_{+,k} + \left.\frac{\partial H}{\partial x}\right|_{-,k} \right) \right] \nonumber\\
    & \quad + \delta x_{-,N}\left[ -\lambda_{+,N} + p_{+,N-1+\frac12} \right] \nonumber\\
    & \quad - \delta \lambda_{+,N}[x_{-,N}] \nonumber\\
    & \quad + \delta x_{+,N}\left[ p_{-,N-1+\frac12} \right] \nonumber\\
    & \quad + \sum_{k=0}^{N-1}\delta p_{+,k+\frac12}\left[ x_{-,k+1}-x_{-,k} - \epsilon \left( \left.\frac{\partial H}{\partial p}\right|_{+,k} - \left.\frac{\partial H}{\partial p}\right|_{-,k} \right) \right] \nonumber\\
    & \quad + \sum_{k=1}^{N-1} \delta x_{+,k}\left[ p_{-,k-1+\frac12}-p_{-,k+\frac12} - \epsilon\left( \left.\frac{\partial H}{\partial x}\right|_{+,k} - \left.\frac{\partial H}{\partial x}\right|_{-,k} \right) \right] \nonumber\\
    & \quad + \delta x_{+,0}\left[ \lambda_{-,0}-p_{-,\frac12} - \epsilon\left( \left.\frac{\partial H}{\partial x}\right|_{+,0} - \left.\frac{\partial H}{\partial x}\right|_{-,0} \right) \right].
\end{align}

Since the variations $\delta x_{\pm,k},\,\delta p_{\pm,k+\frac12},$ and $\delta \lambda_\pm$ are all independent and arbitrary, in order for \cref{eq:discreteHRpmvariation1} to be identically equal to zero, each term inside the various square brackets must be identically equal to zero.  We thus arrive at the discrete, classical Schwinger-Keldysh equations of motion, as well as their initial conditions, in the $+/-$ coordinates, all found from the $\hbar\rightarrow0$ limit of the phase space path integral (\cref{eq:expectationvaluepm1,eq:discreteHRaction}):
\begin{subequations}
    \label{eq:discreteHRpmEOM}
    \begin{align}
    \text{\shortstack{Initial \\[5pt] Conditions}}\quad
    &\left\{
    \begin{aligned}
    \quad\quad\! x_{+,0} &= x_0, \\
    p_{+,\frac12} &= p_0 - \frac\epsilon2
    \left(\left.\frac{\partial H}{\partial x}\right|_{+,0}
        + \left.\frac{\partial H}{\partial x}\right|_{-,0} \right)
    \end{aligned}
    \right.
    \label{eq:ICplus}
    \\[1ex]
    \text{\shortstack{Equations \\[5pt] of Motion}}\quad
    &\left\{
    \begin{aligned}
    \quad x_{+,k+1} &= x_{+,k} + \frac\epsilon2
    \left(\left.\frac{\partial H}{\partial p}\right|_{+,k}
        + \left.\frac{\partial H}{\partial p}\right|_{-,k} \right), \quad k = 0,\dots,N-1,
    \\
    p_{+,k+\frac12} &= p_{+,k-1+\frac12} - \frac\epsilon2
    \left(\left.\frac{\partial H}{\partial x}\right|_{+,k}
        + \left.\frac{\partial H}{\partial x}\right|_{-,k} \right), \quad k = 1,\dots,N-1
    \end{aligned}
    \right.
    \label{eq:EOMplus}
    \\[1ex]
    \text{Lag.\ Mult.\ }\quad
    &
    \qquad\quad\!\lambda_{+,N} = p_{+,N-1+\frac12}
    \label{eq:HamlpN}
    \\[2ex]
    \text{\shortstack{``Initial'' \\[5pt] Conditions}}\quad
    &\left\{
    \begin{aligned}
    x_{-,N} &= 0, \\
    p_{-,N-1+\frac12} &= 0
    \end{aligned}
    \right.
    \label{eq:ICminus}
    \\[1ex]
    \text{\shortstack{Equations \\[5pt] of Motion}}\quad
    &\left\{
    \begin{aligned}
    \;x_{-,k} &= x_{-,k+1}
    - \epsilon\left(\left.\frac{\partial H}{\partial p}\right|_{+,k}
                - \left.\frac{\partial H}{\partial p}\right|_{-,k} \right), \quad k = 0,\dots,N-1,
    \\
    \;p_{-,k-1+\frac12} &= p_{-,k+\frac12}
    + \epsilon\left(\left.\frac{\partial H}{\partial x}\right|_{+,k}
                - \left.\frac{\partial H}{\partial x}\right|_{-,k} \right), \quad k = 1,\dots,N-1
    \end{aligned}
    \right.
    \label{eq:EOMminus}
    \\[1ex]
    \text{Lag.\ Mult.\ }\quad
    & 
    \qquad\quad\; \lambda_{-,0} =
    p_{-,\frac12} + \epsilon\left(
    \left.\frac{\partial H}{\partial x}\right|_{+,0}
    - \left.\frac{\partial H}{\partial x}\right|_{-,0} \right)
    \label{eq:Hamlm0}
    \end{align}
\end{subequations}

The interpretation of the above equations is as follows: \cref{eq:ICplus} give the initial conditions for the $+$ position and momentum coordinates at the initial time slice $k=0$ (where the offset in the initial momentum $p_{+,\frac12}$ from $p_0$ due to a force from the potential $V$ emphasizes that one should really think of the time slicing associated with the discretized momentum as offset from the position time slicing); \cref{eq:EOMplus} give the discrete (forward) time evolution of the $+$ coordinates; \cref{eq:ICminus} give the ``initial'' conditions for the $-$ coordinates at the final time slice $k=N$ (where, again, the discretized momentum is offset from the position time slicing); and \cref{eq:EOMminus} give the discrete (backward) time evolution of the $-$ coordinates.  We emphasize that, fascinatingly, the equations of motion for the $-$ coordinates evolve the $-$ coordinates \emph{backwards} in time.  Finally, \cref{eq:HamlpN,eq:Hamlm0} give the values of the Lagrange multipliers $\lambda_\pm$ at the initial and final time slices; the Lagrange multipliers naturally adjust themselves as necessary such that the equations for the $x_\pm,\,p_\pm$ are fully satisfied for all times; without these Lagrange multipliers, the discrete time evolution would lead to a reduced order of evolution for the coordinates at the final step of the evolution forward (backward) in time for the $+$ ($-$) positions.

We may make progress by seeing that we may \emph{exactly} solve the equations for the $-$ coordinates.  The logic is as follows.  $x_{-,N}=0$ and $p_{-,N-1+\frac12}=0$ from the ``initial'' conditions \cref{eq:ICminus}; if we assume that $x_{-,N-1}=0$, we see that
\begin{align}
    \left.\frac{\partial H}{\partial p}\right|_{+,N-1} & = \; \left.\frac{\partial H}{\partial p}\right|_{-,N-1}.
\end{align}
Then from the dynamics of $x_{-,k}$ of \cref{eq:EOMminus}, we have, self-consistently, that
\begin{align}
    \label{eq:Hamxmk2}
    x_{-,N-1} & = 0.
\end{align}
We are guaranteed the uniqueness of this solution $x_{-,N-1} = 0$ by the Implicit Function Theorem so long as the Hamiltonian is not pathological,
\begin{align}
    \frac{\partial}{\partial x}\frac{\partial H}{\partial p} \ne 1.
\end{align}
Then from the dynamics of $p_{-,k+\frac12}$ of \cref{eq:EOMminus} we have that
\begin{align}
    p_{-,N-2+\frac12} = 0,
\end{align}
since
\begin{align}
    \label{eq:Hampmk2}
    \left.\frac{\partial H}{\partial x}\right|_{+,N-1} = \; \left.\frac{\partial H}{\partial x}\right|_{-,N-1}.
\end{align}
This chain of reasoning may be continued for the $x_{-,k}$ and $p_{-,k}$ coordinates all the way back to $k=0$, yielding
\begin{equation}
    \begin{aligned}
        x_{-,k} & = 0, \qquad k = 0,\dots,N, \\
        p_{-,k+\frac12} & = 0, \qquad k = 0,\dots,N-1.
    \end{aligned}
\end{equation}
One sees that the $-$ coordinates are \emph{dynamically} set to 0 by the equations of motion and the initial conditions determined uniquely by the classical limit $\hbar\rightarrow0$ and the Method of Stationary Phase; we do not need to set the $-$ coordinates to zero by hand \cite{Berges:2007ym}.  

Now that we have the solution for the $-$ coordinates, we may insert these solutions into the equations for the $+$ coordinates, \cref{eq:ICplus,eq:EOMplus}.  Doing so, we find the discrete version of Hamilton's Equations:
\begin{subequations}
    \label{eq:discreteHamiltons}
    \begin{align}
    \text{\shortstack{Initial\\[5pt]Conditions}}\quad
    &\left\{
    \begin{aligned}
    \quad x_{+,0} &= x_0, \\
    p_{+,\frac12} &= p_0 - \epsilon\left.\frac{\partial H}{\partial x}\right|_{0}
    \end{aligned}
    \right.
    \\[2ex]
    \text{\shortstack{Equations\\[5pt]of Motion}}\quad
    &\left\{
    \begin{aligned}
    x_{+,k+1} &= x_{+,k}
    + \epsilon\left.\frac{\partial H}{\partial p}\right|_{k}, \qquad\quad\! k = 0,\dots,N-1, \\
    p_{+,k+\frac12} &= p_{+,k-1+\frac12}
    - \epsilon\left.\frac{\partial H}{\partial x}\right|_{k}, \qquad k = 1,\dots,N-1,
    \end{aligned}
    \right.
    \end{align}
\end{subequations}
where
\begin{align}
    \left.\frac{\partial H}{\partial x}\right|_{k} \equiv \left.\frac{\partial H}{\partial x}\right|_{p = p_{+ ,k+\frac12},\,x = x_{+ ,k}}.
\end{align}

We thus have shown that the critical point of the action $S_{HR}$ exists (by construction), is unique, and yields the classical path through phase space associated with the \emph{initial} conditions $x_0$ and $p_0$ and the potential $V(x)$.

With the values of $x_{\pm,k},\,p_{\pm,k+\frac12}$, and $\lambda_\pm$ at the unique critical point of the Schwinger-Keldysh action $S_{HR}$ determined, we should now investigate the Hessian of $S_{HR}$ in order to complete carrying out the Method of Stationary Phase \cref{eq:stationaryphase} to evaluate the discretized phase space path integral \cref{eq:expectationvaluepm1,eq:discreteHRaction}.  One may show that when considering the mixed partial derivatives taken as $x_{-,0}$, $p_{+,\frac12}$,\ldots, $x_{-,N-1}$, $p_{+,N-1+\frac12}$, $x_{-,N}$, $\lambda_{+,N}$, $x_{+,N}$, $p_{-,N-1+\frac12}$,\ldots, $x_{+,1}$, $p_{-,\frac12}$, $x_{+,0}$, $\lambda_{-,0}$ and when evaluated at the critical point with $x_{-,k}=p_{-,k+\frac12}=\lambda_{-,0}=0$, the Hessian of $S_{HR}$ has the following tridiagonal form,
\begin{align}
    \label{eq:discreteHRhessian}
    \mathrm{Hess}(S_{HR}) = \left( 
        \begin{array}{cccccc}
        0 & -1 & & & & \\
        -1 & 0 & 1 & & & \\
         & 1 & 0 & & & \\
         &  & & \ddots & & \\
         &  & & & 0 & 1 \\
         &  & & & 1 & 0 \\
    \end{array}
        \right) + \mathcal O(\epsilon).
\end{align}
with all other entries 0.  

One has in general for a tridiagonal matrix 
\begin{align}
    A_n \equiv \begin{pmatrix}
        a_1 & b_1 \\
        c_1 & a_2 & b_2 \\
        & c_2 & \ddots & \ddots \\
        & & \ddots & \ddots & b_{n-1} \\
        & & & c_{n-1} & a_n
    \end{pmatrix}
\end{align}
that its determinant $f_n\equiv\det A_n$ is given recursively by the continuant 
\begin{align}
    f_n = a_n \, f_{n-1} - c_{n-1}\,b_{n-1}f_{n-2},
\end{align}
with $f_0 = 1$ and $f_{-1} = 0$ \cite{HornJohnsonMatrixAnalysis}.  

For the case of our Hessian matrix \cref{eq:discreteHRhessian}, then, we have that
\begin{align}
    \det\big(\mathrm{Hess}(S_{HR})\big)  & = (-1)^{4N+4} + \mathcal O (\epsilon)= 1 + \mathcal O (\epsilon).
\end{align}

Let us now turn to the signature of the Hessian matrix.  The characteristic polynomial for the eigenvalues of our tridiagonal matrix is again a tridiagonal matrix.  Using the continuant again, one may see after a few moments of consideration that the characteristic polynomial will always be a polynomial in the eigenvalues $\lambda^2$.  Thus all eigenvalues will come in opposite sign pairs, and the signature will be 0,
\begin{align}
    \sgn\big(\mathrm{Hess}(S_{HR})\big)  & = 0.
\end{align}
Since $\lambda_{-,0}=x_{-,0}=0$ at the critical point of $S_{HR}$, the leftover exponential terms from the initial Gaussian wavepacket in \cref{eq:expectationvaluepm1} safely evaluate to 1, independent of both $\hbar$ and $\sigma_x$, in the limit $\hbar\rightarrow0$.  Finally, at the critical point of the action, one can see that $S_{HR}=0$, and thus there is no overall phase at the end of the evaluation of the Method of Stationary Phase.  
We therefore find, self-consistently, that in the limit $\hbar\rightarrow0$ and $N\rightarrow\infty$ we have exactly the result
\begin{align}
    \label{eq:classicallimitdiscreteHamEqs}
    \langle \hat x \rangle (t_f) = x_{cl}(t_f),
\end{align}
where $x_{cl}(t)$ is the classical solution to Hamilton's Equations for a given Hamiltonian $H$ and with \emph{initial conditions} $x(t_i)=x_0$ and $p(t_i)=p_0$.  Note that in \cref{eq:classicallimitdiscreteHamEqs} all the factors of $2\pi\hbar$ in the denominators of the measure exactly cancel with the factors of $2\pi\hbar$ induced from the integration over the Gaussian fluctuations around the critical point of $S_{HR}$ in the Method of Stationary Phase \cref{eq:stationaryphase}.

\subsection{Continuous Case}
\label{sec:continuousHamilton}

There are several ways in which we may consider the continuum generalization of the discrete phase space path integral.  The most pragmatic is to simply consider the $\epsilon\rightarrow0$ (``continuum extrapolotion'' in the language of lattice quantum field theory) limit of the discrete Hamilton's Equations, \cref{eq:discreteHamiltons}.  One may easily go one step further back and consider the continuum extrapolation of the full set of discrete classical Schwinger-Keldysh equations of motion \cref{eq:discreteHRpmEOM}.  We would, however, like to make contact with the usual continuum variational treatment of the action in classical mechanics \cite{goldstein2002classical}.  We would thus like to consider the continuum limit of the discrete phase space path integral.  Again, we are faced with several choices.  The simplest continuum extrapolation of \cref{eq:expectationvaluepm1} is
\begin{align}
    \langle \hat x \rangle (t_f)
    & = \int\frac{d\lambda_+}{2\pi\hbar}\frac{d\lambda_-}{2\pi\hbar}e^{-\frac{\sigma_x^2\lambda_-^2}{2\hbar^2}}\int \mathcal {D}[x_\pm(t),p_\pm(t)]e^{-\frac{x_-^2(t_i)}{8\sigma_x^2}} 
    e^{\frac{i}{\hbar}\tilde S_{HR}[x_\pm(t),p_\pm(t),\lambda_\pm]}x_+(t_f),
\end{align}
where
\begin{align}
    \label{eq:contHRpmaction1}
    \tilde S_{HR}&[x_\pm(t),p_\pm(t),\lambda_\pm]
     \equiv \nonumber\\
    & \lambda_-\big( x_+(t_i)-x_0 \big) -\lambda_+ x_-(t_f) + p_0 x_-(t_i) \nonumber\\
    & + \int_{t_i}^{t_f}dt \, \Big[ p_+(t)\frac{d}{dt}\big[x_-(t)\big]  + p_-(t)\frac{d}{dt}\big[x_+(t)\big] \\
    & \quad\quad + H\big(p_+(t) - \tfrac12p_-(t),x_+(t) - \tfrac12x_-(t)\big) - H\big(p_+(t) + \tfrac12p_-(t),x_+(t) + \tfrac12x_-(t)\big) \Big]. \nonumber
\end{align}
Notice already a subtlety in the functional formulation of the problem: there is a formally infinite constant $(2\pi\hbar)^{-\infinity}$ hiding in the $\mathcal D[p_\pm(t)]$ functional integration measure.  We will return to this subtlety when we discuss the application of the Method of Stationary Phase in the continuous case.  Notice also that we have replaced the discrete time derivatives of \cref{eq:discreteHRaction} with continuous time derivatives in \cref{eq:contHRpmaction1}.

Despite the aforementioned subtlety regarding the formally infinite constant $(2\pi\hbar)^{-\infinity}$, which will plague any functional path integral formulation, \cref{eq:contHRpmaction1} is unsatisfying.  There is an unequal pairing of variables, with the two discrete Lagrange multipliers left over after functionally integrating over the canonically conjugate $p_\pm$ and $x_\pm$ functions.  We could rectify this imbalance by pulling out from the functional integration two usual integrals over $dx_+(t_i)$ and $dx_-(t_f)$, giving us a dimensionless $\int \frac{d\lambda_+}{2\pi\hbar}\frac{d\lambda_-}{2\pi\hbar}dx_+(t_i)dx_-(t_f)$.  However, we will see that only the ``initial'' condition $x_-(t_f)=0$ guarantees the physical limit $x_-(t)=p_-(t)=0$; by pulling out the $\int dx_-(t_f)$, the ``initial'' condition for $x_-(t)$ in the functional path integral becomes non-zero in general.  Further, the mixing of usual integration and functional integration is unnatural.  

In order to make a more natural phase space path integral formulation, we notice that in the limit $\epsilon\rightarrow0$ \cref{eq:HamlpN,eq:Hamlm0} tell us that the Lagrange multipliers $\lambda_\pm$ are simply the values of the momenta at the temporal endpoints,
\begin{align}
    \begin{aligned}
        \lambda_{+,N} & = p_+(t_f) \\
        \lambda_{-,0} & = p_-(t_i).
    \end{aligned}
\end{align}
A much more natural continuum extrapolation of the discrete phase space path integral to a functional phase space path integral is, then,
\begin{align}
    \label{eq:continuousHRHam}
    \langle \hat x \rangle (t_f)
    & = \int \mathcal {D}[x_\pm(t),p_\pm(t)]e^{-\frac{\sigma_x^2p_-(t_i)^2}{2\hbar^2}}e^{-\frac{x_-^2(t_i)}{8\sigma_x^2}} 
    e^{\frac{i}{\hbar}S_{HR}[x_\pm(t),p_\pm(t)]}x_+(t_f),
\end{align}
where
\begin{align}
    \label{eq:contHRpmaction2}
    S_{HR}&[x_\pm(t),p_\pm(t)] \nonumber\\
    & \equiv 
    p_-(t_i)\big( x_+(t_i)-x_0 \big) - p_+(t_f) x_-(t_f) + p_0 \, x_-(t_i) \nonumber\\
    & \quad\quad + \int_{t_i}^{t_f}dt \, \Big[ p_+(t)\frac{d}{dt}\big[x_-(t)\big] + p_-(t)\frac{d}{dt}\big[x_+(t)\big] \\
    & \quad\quad + H\big(p_+(t) - \tfrac12p_-(t),x_+(t) - \tfrac12x_-(t)\big) - H\big(p_+(t) + \tfrac12p_-(t),x_+(t) + \tfrac12x_-(t)\big) \Big]. \nonumber
\end{align}
One may absorb the temporal endpoint contributions in \cref{eq:contHRpmaction2} into the time integral using Dirac delta functions,
\begin{align}
    \label{eq:contHRpmaction3}
    S_{HR}&[x_\pm(t),p_\pm(t)] \nonumber\\
    & = \int_{t_i}^{t_f}dt \, \Big\{ \big[ p_-(t)\big( x_+(t)-x_0 \big) + p_0 \, x_-(t) \big]\delta(t-t_i) -p_+(t)x_-(t)\delta(t-t_f) \nonumber\\
    & \quad\quad + p_+(t)\frac{d}{dt}\big[x_-(t)\big] + p_-(t)\frac{d}{dt}\big[x_+(t)\big] \\
    & \quad\quad + H\big(p_+(t) - \tfrac12p_-(t),x_+(t) - \tfrac12x_-(t)\big) - H\big(p_+(t) + \tfrac12p_-(t),x_+(t) + \tfrac12x_-(t)\big) \Big\}. \nonumber
\end{align}
By absorbing the temporal endpoints into the time integral these terms can now be considered as sources inside the action integral.  While \cref{eq:contHRpmaction3} is interesting and worth noting, we will find it easier to work with \cref{eq:contHRpmaction2}.  

In order to approach the $\hbar\rightarrow0$ classical limit, we consider the critical point of the now continuous Schwinger-Keldysh action $S_{HR}[x_\pm(t),p_\pm(t)]$ \cref{eq:contHRpmaction2}:
\begin{align}
    \label{eq:contHRpmvariation1}
    \delta S_{HR}[x_\pm(t),p_\pm(t)]
    & = \delta p_-(t_i)\big( x_+(t_i)-x_0 \big) + p_-(t_i)\delta x_+(t_i) \nonumber\\
    & \quad - \delta p_+(t_f) x_-(t_f) - p_+(t_f)\delta x_-(t_f) + p_0\delta x_-(t_i) \nonumber\\
    & \quad + \int_{t_i}^{t_f}dt \, \bigg[ \delta p_+ \frac{d}{dt}\big[x_-\big] + p_+\frac{d}{dt}\big[\delta x_-\big] + \delta p_- \frac{d}{dt}\big[x_+\big] + p_- \frac{d}{dt}\big[\delta x_+\big]\nonumber\\
    & \quad\quad + \left.\frac{\partial H}{\partial p}\right|_{-}\big( \delta p_+ -\frac12\delta p_- \big) + \left.\frac{\partial H}{\partial x}\right|_{-}\big( \delta x_+ -\frac12\delta x_- \big) \nonumber\\
    & \quad\quad - \left.\frac{\partial H}{\partial p}\right|_{+}\big( \delta p_+ +\frac12\delta p_- \big)  - \left.\frac{\partial H}{\partial x}\right|_{+}\big( \delta x_+ +\frac12\delta x_- \big)  \bigg]\nonumber\\
    & = \delta p_-(t_i)\big[ x_+(t_i)-x_0 \big] \nonumber\\
    & \quad + \delta x_-(t_i) \left[ p_+(t_i) - p_0 \right] \nonumber\\
    & \quad + \int_{t_i}^{t_f}dt \, \delta p_-\left[ \frac{d}{dt}\big[x_+\big] - \frac12\left( \left.\frac{\partial H}{\partial p}\right|_{+} + \left.\frac{\partial H}{\partial p}\right|_{-} \right) \right] \nonumber\\
    & \quad - \int_{t_i}^{t_f}dt \, \delta x_-\left[ \frac{d}{dt}\big[p_+\big] + \frac12\left( \left.\frac{\partial H}{\partial x}\right|_{+} + \left.\frac{\partial H}{\partial x}\right|_{-} \right) \right] \nonumber\\
    & \quad - \delta p_-(t_f)\big[ x_-(t_f) \big] \nonumber\\
    & \quad + \delta x_+(t_f) \big[ p_-(t_f) \big] \nonumber\\
    & \quad + \int_{t_i}^{t_f}dt \, \delta p_+\left[ \frac{d}{dt}\big[x_-\big] - \left( \left.\frac{\partial H}{\partial p}\right|_{+} - \left.\frac{\partial H}{\partial p}\right|_{-} \right) \right] \nonumber\\
    & \quad - \int_{t_i}^{t_f}dt \, \delta x_+\left[ \frac{d}{dt}\big[p_-\big] + \left( \left.\frac{\partial H}{\partial x}\right|_{+} - \left.\frac{\partial H}{\partial x}\right|_{-} \right) \right],
\end{align}
where
\begin{align}
    \begin{aligned}
        \left. \frac{\partial H}{\partial p} \right|_{\pm} & \equiv \left. \frac{\partial H}{\partial p} \right|_{\substack{p(t) = p_+(t)\pm\frac12p_-(t), \\[3pt] x(t) = x_+(t)\pm\frac12x_-(t)}}, \\[10pt]
        \left. \frac{\partial H}{\partial x} \right|_{\pm} & \equiv \left. \frac{\partial H}{\partial x} \right|_{\substack{p(t) = p_+(t)\pm\frac12p_-(t), \\[3pt] x(t) = x_+(t)\pm\frac12x_-(t)}}.
    \end{aligned}
\end{align}
Notice that to arrive at the second equality in \cref{eq:contHRpmvariation1} we have integrated by parts, which is the continuum version of the index manipulation we performed when considering the variation of the action for the discrete phase space path integral.  Note, crucially, that we do not constrain any of the $p_\pm$ or $x_\pm$ at the endpoints; i.e.\ we \emph{do not} take $\delta p_\pm(t_i),\,\delta p_\pm(t_f),\,\delta x_\pm(t_i)$ or $\delta x_\pm(t_f)$ to be zero.  It is precisely the contributions from the variations of the phase space variables at the temporal endpoints that give us the correct initial conditions for $p_+(t_i),\,p_-(t_f),\,x_+(t_i)$, and $x_-(t_f)$ as we will explicitly show in a moment.  

Since the variations $\delta x_\pm(t)$ and $\delta p_\pm(t)$ are all independent and arbitrary, in order for \cref{eq:contHRpmvariation1} to be identically equal to zero, each term inside the various square brackets must itself be identically equal to zero.  We thus arrive at the continuous, classical Schwinger-Keldysh equations of motion, as well as their initial conditions, in the $+/-$ coordinates:
\begin{subequations}
    \begin{align}
        \label{eq:contICplus}
    \text{\shortstack{Initial\\[5pt]Conditions}}\quad
    &\left\{
    \begin{aligned}
    \, x_{+}(t_i) &= x_0, \\
    p_{+}(t_i) &= p_0
    \end{aligned}
    \right.
    \\[2ex]
        \label{eq:contEOMplus}
    \text{\shortstack{Equations\\[5pt]of Motion}}\quad
    &\left\{
    \begin{aligned}
    \quad \frac{d x_{+}}{dt}  &= \frac12\left( \left.\frac{\partial H}{\partial p}\right|_{+} + \left.\frac{\partial H}{\partial p}\right|_{-} \right), \\[5pt]
     \frac{d p_{+}}{dt} &= -\frac12\left( \left.\frac{\partial H}{\partial x}\right|_{+} + \left.\frac{\partial H}{\partial x}\right|_{-} \right)
    \end{aligned}
    \right.
    \\[2ex]
        \label{eq:contICminus}
    \text{\shortstack{``Initial''\\[5pt]Conditions}}\quad
    &\left\{
    \begin{aligned}
    \, x_{-}(t_f) &= 0, \\
    p_{-}(t_f) &= 0
    \end{aligned}
    \right.
    \\[2ex]
        \label{eq:contEOMminus}
    \text{\shortstack{Equations\\[5pt]of Motion}}\quad
    &\left\{
    \begin{aligned}
    \quad \frac{d x_{-}}{dt}  &= \left.\frac{\partial H}{\partial p}\right|_{+} - \left.\frac{\partial H}{\partial p}\right|_{-} , \\[5pt]
    \frac{d p_{-}}{dt} &= -\left( \left.\frac{\partial H}{\partial x}\right|_{+} - \left.\frac{\partial H}{\partial x}\right|_{-} \right).
    \end{aligned}\right.
    \end{align}
\end{subequations}
Notice that we have not accidentally missed a sign in the minus coordinates' equations of motion compared to the discrete case; the signs are flipped in the discrete case as the backwards time evolution is made explicit there by solving for the earlier time-sliced coordinates in terms of the later time-sliced coordinates.  Here, in the continuous case, the backwards time evolution is made implicit by the final time ``initial'' conditions, \cref{eq:contICminus}.

Similarly to the discrete case, we may exactly solve the equations of motion for the $-$ coordinates \cref{eq:contEOMminus} subject to the ``initial'' conditions at $t=t_f$, \cref{eq:contICminus}.  In this case, the solution is even easier to see: $x_-(t)=0$ and $p_-(t)=0$ for all times $t\in[t_i,t_f]$ is clearly a solution; by the uniqueness of solutions to the initial value problem in ordinary differential equations guaranteed by the Picard-Lindel\"of Theorem \cite{CoddingtonLevinson1955}, this solution is the only solution.

Again, similarly to the discrete case, we have that $p_-(t_i)=0$ from the equations of motion derived from the Method of Stationary Phase, and we are again able to safely take $\hbar\rightarrow0$ in the $\exp[-\sigma_x^2p_-(t_i)^2/2\hbar^2]$ term; since $x_-(t_i)=0$ similarly, we are also able to safely take $\sigma_x\rightarrow0$ in the remaining $\exp[-x_-^2(t_i)/8\sigma_x^2]$ term.

With $x_-(t)=0$ and $p_-(t)=0$ for all times, \cref{eq:contICplus,eq:contEOMplus} become the usual Hamilton's Equations, as an \emph{initial value problem}:
\begin{subequations}
    \begin{align}
    \text{\shortstack{Initial\\[5pt]Conditions}}\quad
    &\left\{
    \begin{aligned}
        \label{eq:contHamIC}
    \,\,\, x(t_i) &= x_0, \\
    p(t_i) &= p_0
    \end{aligned}
    \right.
    \\[2ex]
    \text{\shortstack{Equations\\[5pt]of Motion}}\quad
    &\left\{
    \begin{aligned}
        \label{eq:contHamEOM}
    \;\, \left. \frac{dx}{dt} \right|_t &= \left.\frac{\partial H}{\partial p}\right|_{\substack{p = p(t) \\[2pt] x = x(t)}}, \\[5pt]
    \left. \frac{dp}{dt} \right|_t &= -\left.\frac{\partial H}{\partial x}\right|_{\substack{p = p(t) \\[2pt] x = x(t)}}.
    \end{aligned}
    \right.
    \end{align}    
\end{subequations}

One would like to complete the consideration of the classical $\hbar\rightarrow0$ limit of the continuous phase space path integral by fully evaluating the result of the Method of Stationary Phase, including taking the determinant of the Hessian of the action.  Perhaps not surprisingly, the evaluation of a functional determinant is significantly more subtle than computing the determinant of a finite-dimensional matrix.  Although we would ideally like to rigorously compute the functional determinant from applying the Method of Stationary Phase to \cref{eq:continuousHRHam}, it turns out that one is not able to do so in general.  Rather, the functional determinant is only generally well-defined as a ratio \cite{Dunne:2007rt}.  To be well-defined in the continuous phase space path integral case, we need to explicitly consider
\begin{align}
    \label{eq:expectationforPI}
    \langle \hat x\rangle(t_f) 
    & = \frac{\langle\psi(t_i)|\hat U^\dagger_{\hat H}(t_f,t_i)\,\hat x \,\hat U_{\hat H}(t_f,t_i)|\psi(t_i)\rangle}{\langle\psi(t_i)|\hat U^\dagger_{\hat H}(t_f,t_i) \,\hat U_{\hat H}(t_f,t_i)|\psi(t_i)\rangle}.
\end{align}
For a normalized initial state and for unitary time evolution such as we consider, the denominator in \cref{eq:expectationforPI} is obviously unity.  However, the introduction of the denominator allows us to rigorously compute the functional derivative from the continuous phase space path integral in the $\hbar\rightarrow0$ limit.  (It is sometimes said that the measure of the functional integral is \emph{defined} by the requirement that the denominator of \cref{eq:expectationforPI} is unity \cite{FeynmanHibbs1965,Schulman:1981vu}.)  In particular, we have that in the $\hbar\rightarrow0$ limit, since $S_{HR}[x_{\pm,\,cl}(t),p_{\pm,\,cl}(t)] = 0$ and $x_{-,\,cl}(t)=p_{-,\,cl}(t)=0$,
\begin{align}
    \langle \hat x\rangle(t_f) 
    & =\frac{\int \mathcal {D}[x_\pm(t),p_\pm(t)]e^{-\frac{\sigma_x^2p_-(t_i)^2}{2\hbar^2}}e^{-\frac{x_-^2(t_i)}{8\sigma_x^2}}
    e^{\frac{i}{\hbar}S_{HR}[x_\pm(t),p_\pm(t)]}x_+(t_f)}{\int \mathcal {D}[x_\pm(t),p_\pm(t)]e^{-\frac{\sigma_x^2p_-(t_i)^2}{2\hbar^2}}e^{-\frac{x_-^2(t_i)}{8\sigma_x^2}} 
    e^{\frac{i}{\hbar}S_{HR}[x_\pm(t),p_\pm(t)]}} \nonumber\\
    & = \frac{x_{cl}(t)\left[ \int\mathcal D[\eta_\pm(t),\pi_\pm(t)]e^{i\frac12\eta^T \big( \Hess(S) \big) \eta} \right]}{\left[\int\mathcal D[\eta_\pm(t),\pi_\pm(t)]e^{i\frac12\eta^T \big( \Hess(S) \big) \eta}\right]} \nonumber\\
    & = x_{cl}(t_f),
\end{align}
where $x_{cl}(t_f)$ solves \cref{eq:contHamIC,eq:contHamEOM}.  While the cancellation between the remaining phase space path integral over the fluctuations in the numerator and denominator obviously cancel, we include for completeness the quantities that define the remaining phase space path integral over the fluctuations:
\begin{align}
    \begin{aligned}
        x_\pm(t) & \equiv x_{\pm,\,cl}(t) + \sqrt{\hbar}\,\eta_\pm(t) \\
        p_\pm(t) & \equiv p_{\pm,\,cl}(t) + \sqrt{\hbar}\,\pi_\pm(t), \\
        \eta & \equiv \left( 
            \begin{array}{c}
                \eta_+(t) \\
                \pi_+(t) \\
                \eta_-(t) \\
                \pi_-(t)
            \end{array}
            \right), \\
        \Hess(S) & = 
            \left(
                \begin{array}{cc}
                    0 & B \\
                    B^T & 0
                \end{array}
            \right), \\
        B & \equiv 
            \left(
                \begin{array}{cc}
                    -\frac{\partial^2 H}{\partial x^2} & \delta(t-t_f)-\partial_t-\frac{\partial^2H}{\partial x\partial p} \\
                    -\delta(t-t_f)+\partial_t-\frac{\partial^2H}{\partial p\partial x} & -\frac{\partial^2 H}{\partial p^2}
                \end{array}
            \right) \\
        B^T & \equiv
            \left(
                \begin{array}{cc} 
                    -\frac{\partial^2 H}{\partial x^2} & -\delta(t-t_i)-\partial_t-\frac{\partial^2H}{\partial x\partial p} \\
                    \delta(t-t_i)+\partial_t-\frac{\partial^2H}{\partial p\partial x} & -\frac{\partial^2 H}{\partial p^2} 
                \end{array}
            \right).
    \end{aligned}
\end{align}
Note that we define the transpose $B^T$ through
\begin{align}
    \begin{aligned}
        \int_{t_i}^{t_f}dt\,\eta_+^T B^T \eta_- & \equiv \int_{t_i}^{t_f}dt\,\eta_-^T B \eta_+, \\
        \eta^T_\pm & \equiv (\eta_\pm \;\; \pi_\pm).
    \end{aligned}
\end{align}

\section{Newton's Second Law from Schwinger-Keldysh}
\subsection{Discrete Case}
\label{sec:discreteNII}
Let us now proceed to the position space path integral from the phase space path integral by integrating out the momenta.  In the usual way, we will rely on the Gaussian integral formula
\begin{align}
    \int_{-\infty}^{\infty}dp \, e^{\pm i\, a \, p^2\mp i\,b\,p} = e^{\pm i \frac{\pi}{4}} \sqrt{\frac{\pi}{a}} e^{\mp i \frac{b^2}{4a}};
\end{align}
therefore we must integrate out the momenta in the $1/2$ coordinates first and only afterwards change to the $+/-$ coordinates.  Starting from \cref{eq:expectationvalue122}, we find the following path integral
\begin{multline}
    \label{eq:naiveNIIPI}
    \langle \hat x \rangle (t_f)
    = \lim_{N\rightarrow\infinity}\left( \frac{m}{2\pi\hbar\epsilon} \right)^N \prod_{n=0}^N \int dx_{+,n}dx_{-,n} x_{+,N} \int\frac{d\lambda_+}{2\pi\hbar}\frac{d\lambda_-}{2\pi\hbar} \\
    e^{-\frac{x_{-,0}^2}{8\sigma_x^2}}e^{-\frac{\sigma_x^2\lambda_-^2}{2\hbar^2}} 
    \exp\left\{\frac i\hbar S_{HR}[x_{\pm,k},\lambda_\pm]\right\},
\end{multline}
where
\begin{align}
    \label{eq:discreteNIIS}
    S_{HR}[x_{\pm,k},\lambda_\pm] & \equiv 
    p_0 x_{-,0} - \lambda_+ x_{-,N} + \lambda_-\big( x_{+,0}-x_0 \big) \nonumber\\
    & \quad + \epsilon\sum_{k=0}^{N-1}\Bigg\{ \frac{m}{\epsilon^2}\left(x_{+,k+1}-x_{+,k}\right)\left(x_{-,k+1}-x_{-,k}\right) \nonumber\\
    & \qquad - V\left(x_{+,k} + \frac12x_{-,k}\right) + V\left(x_{+,k} - \frac12x_{-,k}\right) \Bigg\}.
\end{align}

We may now take the $\hbar\rightarrow0$ limit of \cref{eq:naiveNIIPI} utilizing again the Method of Stationary Phase.  We must thus find the critical point of the action $\delta S_{HR}[x_{\pm,k},\lambda_\pm]$.  Doing so, one finds
\begin{subequations}
    \begin{align}
        \text{\shortstack{Initial \\[5pt] Conditions}}\quad
        &\left\{
        \begin{aligned}
        \quad\,\, x_{+,0} &= x_0 \\
        x_{+,1} &= x_{+,0} + \epsilon\frac{p_0}{m} - \frac{\epsilon^2}{2m}\left(\left.\frac{\partial V}{\partial x}\right|_{+,0} + \left.\frac{\partial V}{\partial x}\right|_{-,0}\right)
        \end{aligned}
        \right.
        \label{eq:NIIICplus}
        \\[1ex]
        \text{EOM}\quad
        & \left\{
        \,\, x_{+,k+1} = 2x_{+,k}-x_{+,k-1}-\frac{\epsilon^2}{2m}\left(\left.\frac{\partial V}{\partial x}\right|_{+,k} + \left.\frac{\partial V}{\partial x}\right|_{-,k}\right),\right. \nonumber\\ 
        & \qquad\qquad\qquad\qquad\qquad\qquad\qquad\qquad\qquad\qquad\qquad k = 1,\ldots,N-1
        \label{eq:NIIEOMplus}
        \\[1ex]
        \text{Lag.\ Mult.\ }\quad
        &
        \qquad\quad\! \lambda_{+} = m\frac{x_{+,N}-x_{+,N-1}}{\epsilon}
        \label{eq:NIIlpN}
        \\[2ex]
        \text{\shortstack{``Initial'' \\[5pt] Conditions}}\quad
        &\left\{
        \begin{aligned}
        x_{-,N} &= 0 \\
        x_{-,N-1} &= x_{-,N}
        \end{aligned}
        \right.
        \label{eq:NIIICminus}
        \\[1ex]
        \text{\shortstack{EOM}}\quad
        & \left\{ \,\, x_{-,k-1} = 2x_{-,k}-x_{-,k+1}-\frac{\epsilon^2}{2m}\left(\left.\frac{\partial V}{\partial x}\right|_{+,k} - \left.\frac{\partial V}{\partial x}\right|_{-,k}\right),\right. \nonumber\\
        & \qquad\qquad\qquad\qquad\qquad\qquad\qquad\qquad\qquad\qquad\qquad k = 1,\ldots,N-1
        \label{eq:NIIEOMminus}
        \\[1ex]
        \text{Lag.\ Mult.\ }\quad
        & 
        \qquad\;\,\,\, \lambda_{-} =
        m\frac{x_{-,1}-x_{-,0}}{\epsilon} + \epsilon\left(\left.\frac{\partial V}{\partial x}\right|_{+,0} - \left.\frac{\partial V}{\partial x}\right|_{-,0}\right).
        \label{eq:NIIlm0}
    \end{align}
\end{subequations}
As before, one immediately sees that the necessary and unique solution for the minus coordinates driven by the $\hbar\rightarrow0$ limit is $x_{-,k}=0$ for all $k$, where, again, the solution propagates backwards from the final time slice.  The initial conditions and the equations of motion for the plus coordinates then collapse to the discrete version of Newton's Second Law as an \emph{initial value problem}:
\begin{subequations}
    \begin{align}
        \text{\shortstack{Initial \\[5pt] Conditions}}\quad
        &\left\{
        \begin{aligned}
        \quad\! x_{+,0} &= x_0 \\
        x_{+,1} &= x_{+,0} + \epsilon\frac{p_0}{m} - \frac{\epsilon^2}{m}\left.\frac{\partial V}{\partial x}\right|_{0}
        \end{aligned}
        \right.
        \label{eq:NIIICplus2}
        \\[1ex]
        \text{EOM}\quad
        &
        \quad x_{+,k+1} = 2x_{+,k}-x_{+,k-1}-\frac{\epsilon^2}{m}\left.\frac{\partial V}{\partial x}\right|_{+,k},\qquad k = 1,\ldots,N-1
        \label{eq:NIIEOMplus2}
        \\[1ex]
        \text{Lag.\ Mult.\ }\quad
        &
        \qquad\,\,\,\lambda_{+} = m\frac{x_{+,N}-x_{+,N-1}}{\epsilon}.
        \label{eq:NIIlpN2}
    \end{align}
\end{subequations}
One fascinating aspect of \cref{eq:NIIEOMplus2,eq:NIIlpN2} is how the discrete second derivative persists through the final time slice.  One might have expected that the dynamics would lose one order at the final time slice, as commonly occurs when discretizing continuous equations of motion.  What we see here is that the Lagrange multiplier $\lambda_+$ adjusts as necessary in order to maintain the fully correct discrete second derivative.  (One can see a similar story for the minus coordinates in \cref{eq:NIIICminus,eq:NIIlm0}.)  

Unfortunately it is not possible to put the Hessian into a tridiagonal form as was done in the discrete phase space case as there exists a cycle amongst the coordinates other than $\lambda_-$ and $x_{+,N}$ \cite{brouwer2011spectra}.  One may show that the Hessian for \cref{eq:discreteNIIS}, when the mixed partials are evaluated in the order $\lambda_+,\,,x_{+,N},\ldots,x_{+,0},x_{-,N},\ldots,x_{-,0},\lambda_-$, is in the block diagonal form
\begin{align}
    \Hess(S_{HR}) = \left( 
        \begin{array}{cc}
            0 & B \\
            B^T & 0
        \end{array}
    \right),
\end{align}
where
\begin{align}
    B = 
    \left(
        \begin{array}{cccccc}
            -1 & 0 &&&& \\
            \frac{m}{\epsilon} & -\frac{m}{\epsilon} & 0 &&& \\
            -\frac{m}{\epsilon} & \frac{2m}{\epsilon} & -\frac{m}{\epsilon} & 0 && \\
            0 & -\frac{m}{\epsilon} & \frac{2m}{\epsilon} & -\frac{m}{\epsilon} & 0 & \\
            &&&\ddots&& \\
            & 0 & -\frac{m}{\epsilon} & \frac{2m}{\epsilon} & -\frac{m}{\epsilon} & 0 \\
            & & 0 & -\frac{m}{\epsilon} & \frac{m}{\epsilon} & 1
        \end{array}
    \right) + \mathcal O(\epsilon),
\end{align}
and all other entries are 0.  This block form for the Hessian implies that, to leading order in $\epsilon$,
\begin{align}
    |\det[\Hess(S_{HR})]| 
    & = \left( \frac{m}{\epsilon} \right)^{2N}.
\end{align}
and
\begin{align}
    \sgn[\Hess(S_{HR})] = 0.
\end{align}

Therefore the Method of Stationary Phase \cref{eq:stationaryphase} yields in the $\hbar\rightarrow0$ and $N\rightarrow\infinity$ limit
\begin{align}
    \langle \hat x \rangle (t_f)
    & = \left(\frac{m}{2\pi\hbar\epsilon}\right)^N\left(\frac{1}{2\pi\hbar}\right)^2\left[\left(\frac{m}{\epsilon}\right)^{2N}\right]^{-1/2}\left(2\pi\hbar\right)^{N+2}x_{cl}(t) \nonumber\\
    & = x_{cl}(t_f),
\end{align}
where $x_{cl}(t)$ solves the equations of motion from Newton's Second Law subject to the initial conditions $x(t_i)=x_0$ and $\left.\frac{dx}{dt}\right|_{t=t_i}=v_0,$ $v_0\equiv p_0/m$.

\subsection{Continuous Case}
\label{sec:continuousNII}
We would now like to take the $N\rightarrow\infinity$ limit to consider the case when the particle's path may be considered continuous.  We again have many options for how we may choose to take this limit, e.g.\ at the level of the initial conditions and the equations of motion or earlier in the treatment.  In order to arrive at the most natural treatment of the problem, we will start from the position space path integral in the discrete case \cref{eq:naiveNIIPI} with action given by \cref{eq:discreteNIIS}.  We will first integrate out the $\lambda_+$ degree of freedom, returning the original delta function at the final time slice $\delta(x_{-,N})$.  We may then collapse this delta function by integrating over $x_{-,N}$ and setting any instances of $x_{-,N}$ in the integrand to 0.  We would like to perform a similar procedure for the $x_{+,0}$ coordinate.  First, we integrate out the $\lambda_-$ degree of freedom, yielding
\begin{multline}
    \label{eq:naiveNIIPIa}
    \langle \hat x \rangle (t_f)
    = \lim_{N\rightarrow\infinity}\left( \frac{m}{2\pi\hbar\epsilon} \right)^N \int dx_{+,N} x_{+,N} \prod_{n=0}^{N-1} \int dx_{+,n}dx_{-,n} \\
    \left( \frac{1}{2\pi\sigma_x^2} \right)^{1/2} e^{-\frac{(x_{+,0}-x_0)^2}{2\sigma_x^2}}e^{-\frac{x_{-,0}^2}{8\sigma_x^2}} 
    \exp\left\{\frac i\hbar S_{HR}[x_{\pm,k}]\right\},
\end{multline}
where
\begin{align}
    \label{eq:discreteNIISa}
    S_{HR}[x_{\pm,k}] & \equiv p_0 x_{-,0} + \epsilon\sum_{k=0}^{N-1}\Bigg\{ \frac{m}{\epsilon^2}\left(x_{+,k+1}-x_{+,k}\right)\left(x_{-,k+1}-x_{-,k}\right) \nonumber\\
    & \qquad\qquad\qquad\qquad\qquad - V\left(x_{+,k} + \frac12x_{-,k}\right) + V\left(x_{+,k} - \frac12x_{-,k}\right) \Bigg\}.
\end{align}
The prefactor $(2\pi\sigma_x^2)^{-1/2}\exp[-(x_{+,0}-x_0)^2/(2\sigma_x^2)]$ of the phase in \cref{eq:naiveNIIPIa} is a Dirac delta function in the limit $\sigma_x\rightarrow0$.  However, as noted in \cref{sec:discreteham}, we can only take $\sigma_x\rightarrow0$ \emph{after} we take $\hbar\rightarrow0$.  In order to proceed, we need to perform an asymptotic analysis of the integral over $x_{+,0}$, keeping track of the higher orders in $\sigma_x$.  Such an analysis can be accomplieshed with a straightforward application of Laplace's Method \cite{Wong2001}, in which one expands the potential terms to second order and explicitly integrates out the Gaussian in $x_{+,0}$.  One may then expand the result order by order in $\sigma_x$.  It turns out that the lowest order correction terms in $\sigma_x$ are proportional to $\epsilon$, 
\begin{align}
    \nonumber
    \epsilon\sigma_x\frac{\partial V}{\partial x} \quad \text{and} \quad \epsilon\sigma^2_x\frac{\partial^2 V}{\partial x^2},
\end{align}
and thus for any well-behaved potential these correction terms are zero in the $N\rightarrow\infinity$ limit; one does not even need to consider the $\sigma_x\rightarrow0$ limit after the $\hbar\rightarrow0$ limit.  The result is that the prefactor is actually equivalent to a Dirac delta function centered at $x_0$.  We therefore have that
\begin{align}
    \label{eq:discreteNIIPIb}
    \langle \hat x \rangle (t_f)
    = \lim_{N\rightarrow\infinity}\left( \frac{m}{2\pi\hbar\epsilon} \right)^N \int dx_{+,N} dx_{-,0} x_{+,N}e^{-\frac{x_{-,0}^2}{8\sigma_x^2}}  \prod_{n=1}^{N-1} \int dx_{+,n}dx_{-,n}
    e^{\frac i\hbar S_{HR}[x_{\pm,k}]},
\end{align}
where
\begin{align}
    \label{eq:discreteNIISb}
    S_{HR}[x_{\pm,k}] & \equiv p_0 x_{-,0} + \epsilon\sum_{k=0}^{N-1}\Bigg\{ \frac{m}{\epsilon^2}\left(x_{+,k+1}-x_{+,k}\right)\left(x_{-,k+1}-x_{-,k}\right) \nonumber\\
    & \qquad\qquad\qquad\qquad\qquad - V\left(x_{+,k} + \frac12x_{-,k}\right) + V\left(x_{+,k} - \frac12x_{-,k}\right) \Bigg\},
\end{align}
with $x_{+,0}=x_0$ and $x_{-,N}=0$.  

We now take the formal $N\rightarrow\infinity$ limit to arrive at the continuous position space path integral
\begin{align}
    \label{eq:continuousNII}
    \langle \hat x \rangle (t_f) = \int\mathcal D[x_{\pm}(t)]e^{\frac i\hbar S_{HR}[x_\pm(t)]},
\end{align}
where
\begin{multline}
    \label{eq:continuousNIIaction}
    S_{HR}[x_\pm(t)] = p_0x_-(t_i) \\ + \int_{t_i}^{t_f}dt \, m\frac{d}{dt}\big[ x_+(t) \big]\frac{d}{dt}\big[ x_-(t) \big] - V\Big( x_+(t) + \frac12 x_-(t)\Big) + V\Big( x_+(t) - \frac12 x_-(t)\Big).
\end{multline}
It is clear from \cref{eq:discreteNIIPIb} that the measure in \cref{eq:continuousNII} \emph{does not} include integration over $x_+(t_i)$ or $x_-(t_f)$; those two values are fixed at $x_+(t_i) = x_0$ and $x_-(t_f)=0$.  Thus when performing the Method of Stationary Phase and we consider the variation of the action \cref{eq:continuousNIIaction}, we must remember to set $\delta x_+(t_i) = \delta x_-(t_f) = 0$.  Note, importantly, that no other variations are set to zero; in particular, $\delta x_-(t_i)\ne0$.  We will return to this note when discussing the connection with Galley \cite{Galley:2012hx} in \cref{sec:galley}.  

The variation of the action \cref{eq:continuousNIIaction} then yields
\begin{align}
    \delta S_{HR}[x_\pm(t)]
    & = p_0\delta x_-(t_i) + \int_{t_i}^{t_f}dt\, m\frac{d}{dt}\big[ x_-(t) \big]\frac{d}{dt}\big[ \delta x_+(t) \big] + m\frac{d}{dt}\big[ x_+(t) \big]\frac{d}{dt}\big[ \delta x_-(t) \big] \nonumber\\
    & \qquad - \left. \frac{\partial V}{\partial x} \right|_+\big( \delta x_+(t) + \frac12\delta x_-(t)\big) + \left. \frac{\partial V}{\partial x} \right|_-\big( \delta x_+(t) - \frac12\delta x_-(t)\big) \nonumber\\
    & = \delta x_-(t_i)\left[ p_0 - m\left.\frac{d}{dt} x_+ \right|_{t_i} \right] \nonumber\\
    & \qquad + \delta x_+(t_f)\left[ m\left.\frac{d}{dt}x_-\right|_{t_f}\right] \nonumber\\
    & \qquad + \int_{t_i}^{t_f} dt \, \delta x_-(t)\left[ -m\frac{d^2}{dt^2}x_+ - \frac12\left( \left. \frac{\partial V}{\partial x} \right|_+ + \left. \frac{\partial V}{\partial x} \right|_- \right)\right] \nonumber\\
    & \qquad\qquad + \delta x_+(t) \left[ -m\frac{d^2}{dt^2}x_- + \left. \frac{\partial V}{\partial x} \right|_+ - \left. \frac{\partial V}{\partial x} \right|_- \right],
\end{align}
where
\begin{align}
    \left. \frac{\partial V}{\partial x} \right|_\pm \equiv \left. \frac{\partial V}{\partial x} \right|_{x(t) = x_+(t)\pm\frac12x_-(t)}
\end{align}
and, after the integration by parts, we utilized $\delta x_+(t_i) = \delta x_-(t_f) = 0$.

With $x_+(t_i)=x_0$ and $x_-(t_f)=0$ already fixed, setting the variation of the action to zero from the application of the Method of Stationary Phase yields:
\begin{subequations}
    \begin{align}
        \text{\shortstack{Initial \\[5pt] Conditions}}\quad
        &\left\{
        \begin{aligned}
        \quad\! x_{+}(t_i) &= x_0 \\
        \left.\frac{dx_+}{dt}\right|_{t_i} &= \frac{p_0}{m}
        \end{aligned}
        \right.
        \label{eq:NIIICcplus}
        \\[1ex]
        \text{EOM}\quad
        & \,\,\, m\left.\frac{d^2x_+}{dt^2}\right|_t = -\frac12\left( \left. \frac{\partial V}{\partial x} \right|_+ + \left. \frac{\partial V}{\partial x} \right|_- \right)
        \label{eq:NIIEOMcplus}
        \\[1ex]
        \text{\shortstack{``Initial'' \\[5pt] Conditions}}\quad
        &\left\{
        \begin{aligned}
         \,\,\,\, x_{-}(t_f) &= 0 \\
        \left.\frac{dx_{-}}{dt}\right|_{t_i} &= 0
        \end{aligned}
        \right.
        \label{eq:NIIICcminus}
        \\[1ex]
        \text{\shortstack{EOM}}\quad
        & \,\,\, m\left.\frac{d^2x_-}{dt^2}\right|_t = -\left( \left. \frac{\partial V}{\partial x} \right|_+ - \left. \frac{\partial V}{\partial x} \right|_- \right).
        \label{eq:NIIEOMcminus}
    \end{align}
\end{subequations}

One may see by inspection that $x_-(t)=0$ solves the minus equations of motion \cref{eq:NIIEOMcminus} subject to the initial conditions \cref{eq:NIIICcminus}.  By the Picard-Lindel\"of Theorem \cite{CoddingtonLevinson1955}, this solution is the only solution.  

Dropping the plus subscript for clarity and for a more straightforward connection to our usual classical mechanics result, we are left with the usual Newton's Second Law as an \emph{initial value problem}:
\begin{subequations}
    \begin{align}
    \text{\shortstack{Initial \\[5pt] Conditions}}\quad
    &\left\{
    \begin{aligned}
    \quad\! x(t_i) &= x_0 \\
    \left.\frac{dx}{dt}\right|_{t_i} &= \frac{p_0}{m}
    \end{aligned}
    \right.
    \label{eq:NIIICc}
    \\[1ex]
    \text{EOM}\quad
    & \,\,\, m\left.\frac{d^2x}{dt^2}\right|_t = -\left.\frac{\partial V}{\partial x}\right|_{x(t)}.
    \label{eq:NIIEOMc}
    \end{align}
\end{subequations}

We would naturally next like to fully consider $\langle\hat x\rangle(t_f)$ from the Method of Stationary Phase.  Since $x_-(t_i)$ is set to zero from the variation of the action, we may safely take $\sigma_x\rightarrow0$.  One may examine the Hessian of the action, but, as we saw in the phase space path integral case, the contribution from the second variation of the action in the numerator and denominator will exactly cancel, yielding in the $\hbar\rightarrow0$ limit the expected result
\begin{align}
    \langle \hat x \rangle (t_f) = x_{cl}(t_f),
\end{align}
where $x_{cl}(t)$ solves Newton's Second Law subject to the initial conditions $x_{cl}(t_i)=x_0$ and $d x_{cl}/dt|_{t_i} = p_0/m$.

\section{Euler-Lagrange Equations from Schwinger-Keldysh}

\subsection{Continuous Case: Position Space Path Integral}
\label{sec:continuouspositionPI}
In \cref{sec:discreteNII}, one can see that it is simply impossible to directly and easily connect the discretized coordinate space path integral action that depends only on the discretized positions $x_{k}$ to a discretized path integral action involving a Lagrangian that depends on both $x_k$ and its time derivative.  We will show in \cref{sec:discreteEL} how to make this connection in the discretized case through the introduction of additional Lagrange multipliers.

However, when we go to the continuum limit, there \emph{is} a way to directly and insightfully connect the continuous position space path integral action of \cref{sec:continuousNII} to an action in terms of a Lagrangian.  It should be clear that for our original classical Lagrangian $\frac12m\dot x^2-V(x)$ that the continuum position space path integral \cref{eq:continuousNII} and action \cref{eq:continuousNIIaction} can be written as
\begin{align}
    \label{eq:continuumEL1}
    \langle \hat x\rangle(t_f) = \int \mathcal D[x_\pm(t)]e^{-\frac{x_{-}^2(t_i)}{8\sigma_x^2}}e^{\frac{i}{\hbar}S_{HR}\left[x_\pm(t),\frac{d}{dt}\big[ x_\pm(t) \big]\right]},
\end{align}
where
\begin{multline}
    \label{eq:continuumS1}
    S_{HR}\left[x_\pm(t),\frac{d}{dt}\big[ x_\pm(t) \big] \right] \equiv p_0 x_-(t_i) + \int_{t_i}^{t_f} dt \, \left\{ L\left(x_+(t)+\frac12x_-(t),\frac{d}{dt}\big[ x_+(t)+\frac12x_-(t) \big]\right) \right. \\ \left. - L\left(x_+(t)-\frac12x_-(t),\frac{d}{dt}\big[ x_+(t)+\frac12x_-(t) \big]\right)\right\}.
\end{multline}
Notice, crucially, that the position space path integral \cref{eq:continuumEL1} is over paths $x_\pm(t)$: the position space path integral \emph{only} integrates over the variations of the $x_\pm(t)$, $\delta x_\pm(t)$; there are \emph{no} variations in $\dot x_\pm(t)$, $\delta\dot x_\pm(t)$ as there are no $\dot x_\pm(t)$ in the functional integration measure.  We will consider the full continuum configuration space path integral, in which one has variations in $\dot x_\pm(t)$ as well as in $x_\pm(t)$ in \cref{sec:continuousEL}.  
As in \cref{sec:continuousNII}, the measure in \cref{eq:continuumEL1} does \emph{not} include integrations over $x_+(t_i)$ or $x_-(t_f)$ as those two values for the path are fixed to $x_0$ and $0$, respectively.  

Since there are no variations in $\dot x_\pm(t)$, we completely avoid the issue of the transposition rule.  (Recall that the transposition rule compares $\delta\dot x(t)$ and $\frac{d}{dt}[\delta x(t)]$.)  The minor technical complication is that the variation of the action induced by the Method of Stationary Phase now involves the functional chain rule.  We have in general that
\begin{multline}
    \label{eq:functionalchainrule}
    L\left( x(t) + \delta x(t),\frac{d}{dt}\big[ x(t) + \delta x(t) \big] \right) \\ = L\left( x(t), \frac{d}{dt}\big[ x(t) \big]\right) + \frac{\partial L}{\partial x}\delta x(t) + \frac{\partial L}{\partial \left(\frac{d}{dt}\big[x(t)\big]\right)}\frac{d}{dt}\big[\delta x(t)\big] + \mathcal O\big(\delta x^2(t)\big).
\end{multline}
\cref{eq:functionalchainrule} looks far more complicated that the usual expression involving $\dot x(t)$; however, while notationally heavy, \cref{eq:functionalchainrule} is completely unambiguous and in a form that can be trivially integrated by parts when we consider the variation of the action\footnote{Note that in many treatments, e.g.\ \cite{Flannery2011DAlembertLagrange}, $\delta$ is treated as an operator that may or may not be passed through $d/dt$.  As noted in \cref{sec:notation}, we use $\delta$ in conjunction with a variable, such as $x$, to denote the single variation away from the extremum, $\big(\delta x\big)(t)$, which allows for a change of functional integration measure; e.g., $x_\pm(t) = x_{\pm,cl}(t) + \delta x_\pm(t)$.}.  

Considering the variation of the action and setting it equal to zero, we find the following equations:
\begin{subequations}
    \label{eq:continuumEOMEL}
    \begin{align}
        \text{\shortstack{Initial \\[5pt] Conditions}}\quad
        &\left\{
        \begin{aligned}
            x_+(t_i) & = x_0 \\
           \qquad\qquad\qquad \frac12\left( \left.\frac{\partial L}{\partial\left(\frac{d}{dt}\big[x(t)\big]\right)}\right|_{+,i} + \left.\frac{\partial L}{\partial\left(\frac{d}{dt}\big[x(t)\big]\right)}\right|_{-,i} \right) & = p_0
        \end{aligned}
        \right. \\[1ex]
        \text{\shortstack{Equation \\[5pt] of Motion}}\quad
        &   
            \left. \frac{\partial L}{\partial x} \right|_+ + \left. \frac{\partial L}{\partial x} \right|_- - \frac{d}{dt}\left( \left. \frac{\partial L}{\partial\left(\frac{d}{dt}\big[x(t)\big]\right)} \right|_+ + \left. \frac{\partial L}{\partial\left(\frac{d}{dt}\big[x(t)\big]\right)} \right|_- \right) = 0 \\[1ex]
        \text{\shortstack{``Initial'' \\[5pt] Conditions}}\quad
        &\left\{
        \begin{aligned}
            \label{eq:continuumEL1ICm}
            x_-(t_f) & = 0 \\
            \qquad\qquad\qquad\qquad\, \left. \frac{\partial L}{\partial\left(\frac{d}{dt}\big[x(t)\big]\right)} \right|_{+,f} - \left. \frac{\partial L}{\partial\left(\frac{d}{dt}\big[x(t)\big]\right)} \right|_{-,f} & = 0
        \end{aligned}
        \right. \\
        \label{eq:continuumELEOMm}
        \text{\shortstack{Equation \\[5pt] of Motion}}\quad
        &   \left. \frac{\partial L}{\partial x} \right|_+ - \left. \frac{\partial L}{\partial x} \right|_- - \frac{d}{dt}\left( \left. \frac{\partial L}{\partial\left(\frac{d}{dt}\big[x(t)\big]\right)} \right|_+ - \left. \frac{\partial L}{\partial\left(\frac{d}{dt}\big[x(t)\big]\right)} \right|_- \right) = 0, 
    \end{align}
\end{subequations}
where
\begin{align}
    \left. \frac{\partial L}{\partial\left(\frac{d}{dt}\big[x(t)\big]\right)} \right|_\pm \equiv \, \left. \frac{\partial L}{\partial\left(\frac{d}{dt}\big[x(t)\big]\right)} \right|_{x(t) = x_+(t)\pm\frac12x_-(t)}
\end{align}
and similarly for the other structures in \cref{eq:continuumEOMEL}, with $i$ ($f$) indicating that the function is evaluated at the initial (final) time $t_i$ ($t_f$).  

It is clear from \cref{eq:continuumEL1ICm} that at the final time one has, in the usual way,
\begin{align}
    \begin{aligned}
        x_-(t_f) & = 0 \\
        \left. \frac{d}{dt}\big[ x(t) \big] \right|_{t=t_f} & = 0,
    \end{aligned}
\end{align}
and from the equations of motion \cref{eq:continuumELEOMm} that
\begin{align}
    x_-(t) = 0.
\end{align}
One may then safely take $\sigma_x\rightarrow0$ in the real part of the position space path integral integrand.  We are then left with, dropping the $+$ for notational simplicity and for a more direct connection to the usual formulation of classical mechanics, the Euler-Lagrange equations as an \emph{initial value problem}:
\begin{subequations}
    \begin{align}
    \text{\shortstack{Initial \\[5pt] Conditions}}\quad
    &\left\{
    \begin{aligned}
    x(t_i) &= x_0 \\
    \quad\! \left. \frac{\partial L}{\partial\left(\frac{d}{dt}\big[x(t)\big]\right)} \right|_{t=t_i}  &= p_0
    \end{aligned}
    \right.
    \label{eq:ELICc}
    \\[1ex]
    \text{EOM}\quad
    & \frac{\partial L}{\partial x} - \frac{d}{dt}\frac{\partial L}{\partial\left(\frac{d}{dt}\big[x(t)\big]\right)} = 0.
    \label{eq:ELEOMc}
    \end{align}
\end{subequations}
Again, in the usual way, one has that
\begin{align}
    \langle\hat x\rangle(t_f) = x_{cl}(t_f),
\end{align}
where $x_{cl}(t_f)$ is the classical solution to the Euler-Lagrange equations \cref{eq:ELEOMc} subject to the initial conditions \cref{eq:ELICc}.

\subsection{Discrete Case}
\label{sec:discreteEL}
In order to arrive at a discrete Euler-Lagrange equation, we will need to connect the discrete differences $(x_{\pm,n+1}-x_{\pm,n})/\epsilon$ to the symbol we will associate with time derivatives of the particle's path at discrete instances of time, $\dot x_{\pm,n+\frac12}$.  We will make this connection by first setting the quantities equal through Dirac delta functions.  We will then write these Dirac delta functions as integrals over Lagrange multipliers that will have units of momentum.  By introducing these many Lagrange multipliers, we will arrive at a very natural configuration space path integral with equal numbers of integrals over $x_{\pm,n}$ and $\lambda_{\pm,n}$ that is amenable to the Method of Stationary Phase.  (Should we have kept the path integral only in terms of integrals over $x_\pm(t)$ and $\dot x_\pm(t)$ with the Dirac delta functions connecting the two, we could not have rigorously applied the Method of Stationary Phase.)

Starting from \cref{eq:naiveNIIPI,eq:discreteNIIS}, we insert a very complicated 1:
\begin{align}
    1 & = \prod_{n=1}^{N-1} \int d(\epsilon \dot x_{+,n+\frac12})\delta\big(\epsilon\dot x_{+,n+\frac12}-(x_{+,n+1}-x_{+,n})\big) \nonumber\\
    & \qquad \times \int d(\epsilon \dot x_{-,n+\frac12})\delta\big(\epsilon\dot x_{-,n+\frac12}-(x_{-,n+1}-x_{-,n})\big) \nonumber\\
    & = \epsilon^{2N}\prod_{n=1}^{N-1} d \dot x_{+,n+\frac12}d \dot x_{-,n+\frac12} \frac{d\lambda_{+,n}}{2\pi\hbar}\frac{d\lambda_{-,n+1}}{2\pi\hbar} \nonumber\\
    & \quad \exp\left\{ \frac i\hbar \epsilon\sum_{k=0}^{N-1} \left[ \lambda_{+,k}\left( \frac{x_{-,k+1}-x_{-,k}}{\epsilon} - \dot x_{-,k+\frac12}\right) + \lambda_{-,k+1}\left( \frac{x_{+,k+1}-x_{+,k}}{\epsilon} - \dot x_{+,k+\frac12} \right)\right]\right\}.
\end{align}
Notice the slight shift $k\rightarrow k+1$ between the iterator on the $\lambda_+$ and the $\lambda_-$.  This shift is included in order to include the original $\lambda_{-,0}$, used to enforce the initial wavepacket position, and the $\lambda_{+,N}$, used to enforce the final time slice meeting of the forward and backward paths, in one natural set of $2N$ total Lagrange multipliers.  We could have inserted delta functions of arguments of $\dot x_{+,n+\frac12}-\frac{x_{+,n+1}-x_{+,n}}{\epsilon}$.  However, do so would have led to Lagrange multipliers with units of momentum over time, which would be unnatural.

\cref{eq:naiveNIIPI,eq:discreteNIIS} then becomes
\begin{align}
    \label{eq:discreteEL1}
    \langle \hat x\rangle (t)
    & = \lim_{N\rightarrow\infinity} \prod_{n=0}^N\int dx_{+,n}dx_{-,n}\frac{d\lambda_{+,n}}{2\pi\hbar}\frac{d\lambda_{-,n}}{2\pi\hbar}e^{-\frac{x_{-,0}^2}{8\sigma_x^2}}e^{-\frac{\sigma_x^2\lambda_{-,0}^2}{2}} x_{+,N} \nonumber\\
    & \qquad \prod_{\ell=0}^{N-1}\left( \frac{m\epsilon}{2\pi\hbar} \right)^{1/2}d\dot x_{+,\ell+\frac12}\left( \frac{m\epsilon}{2\pi\hbar} \right)^{1/2}d\dot x_{-,\ell+\frac12}e^{\frac i\hbar S_{HR}[x_{\pm,k},\dot x_{\pm,k+\frac12},\lambda_{\pm,k}]},
\end{align}
where
\begin{align}
    \label{eq:discreteELaction}
    S_{HR}[x_{\pm,k},\dot x_{\pm,k+\frac12},\lambda_{\pm,k}]
    & = p_0 x_{-,0} - \lambda_{+,N}x_{-,N} + \lambda_{-,0}(x_{+,0}-x_0) \nonumber\\
    & \qquad +\epsilon\sum_{k=0}^{N-1}\big\{ 
        L[x_{+,k} + \frac12x_{-,k},\dot x_{+,k+\frac12} + \frac12 \dot x_{-,k+\frac12}] \nonumber\\
    & \qquad\quad   - L[x_{+,k} - \frac12x_{-,k},\dot x_{+,k+\frac12} - \frac12 \dot x_{-,k+\frac12}] \nonumber\\
    & \qquad\quad   +\lambda_{+,k}\left( \frac{x_{-,k+1}-x_{-,k}}{\epsilon} - \dot x_{-,k+\frac12}\right) \nonumber\\
    & \qquad\quad   + \lambda_{-,k+1}\left( \frac{x_{+,k+1}-x_{+,k}}{\epsilon} - \dot x_{+,k+\frac12} \right) \big\}.
\end{align}
Notice that by matching an equal number of Lagrange multipliers (each with units of  momentum) with coordinates, the dimensions work out in \cref{eq:discreteEL1} correctly by inspection.  We now have a rather strange looking path integral in that there are a triplet of coordinates, $x_{\pm,n}$, $\lambda_{\pm,n}$, and $\dot x_{\pm,n+\frac12}$, rather than a pairing of coordinates as we have seen in the past.  Recall, however, that the quantities we are integrating over are not operators (we are integrating over numbers), and so there is no requirement that quantities pair up in canonically conjugate pairs.  

We're interested in the classical $\hbar\rightarrow0$ limit of the path integral \cref{eq:discreteEL1,eq:discreteELaction}.  The measure of the path integral in \cref{eq:discreteEL1} makes it completely transparent what variables the action will be varied with respect to: $x_{\pm,k},\dot x_{\pm,k+\frac12},\lambda_{\pm,k}$.  In particular, the symbols connected to the discrete velocity, $\dot x_{\pm,k+\frac12}$, are allowed to vary \emph{independently} of the coordinates $x_{\pm,k}$, where the connection between those two quantities is enforced through the variation of the Lagrange multipliers $\lambda_{\pm,k}$ (other than the variation of $\lambda_{-,0}$, which will fix the initial position of the particle, and the variation of $\lambda_{+,N}$, which will force the forward and backwards paths to meet at the final time slice).

Variation of the action \cref{eq:discreteELaction} with respect to $x_{\pm,k},\dot x_{\pm,k+\frac12}$, and $\lambda_{\pm,k}$ leads to the now usual situation in which the ``initial conditions'' and equations of motion for the minus coordinates are solved uniquely by $x_{-,k}=\dot x_{-,k+\frac12}=\lambda_{-,k}=0$.  The initial conditions and equations of motion for the plus coordinates then become:
\begin{subequations}
    \begin{align}
        \text{\shortstack{Initial \\[5pt] Conditions}}\quad
        &\left\{
        \begin{aligned}
            x_{+,0} & = x_0 \\
            \label{eq:discreteELlamIC}
            \lambda_{+,0} & = p_0 + \epsilon \left.\frac{\partial L}{\partial x}\right|_{0} \\
            \left.\frac{\partial L}{\partial \dot x}\right|_{0} & = \lambda_{+,0}
        \end{aligned}
        \right.
        \\[1ex]
        \text{First Evolution} \quad & 
            \label{eq:discreteELx1}
            x_{+,1} = x_{+,0} + \epsilon \dot x_{+,\frac12} \\[1ex]
        \text{\shortstack{Equations \\[5pt] of Motion}}\quad
        &\left\{
        \begin{aligned}
            \lambda_{+,k} & = \lambda_{+,k-1} + \epsilon \left.\frac{\partial L}{\partial x}\right|_{k}, & k = 1,\ldots,N-1 \\
            \label{eq:discreteELxdEOM}
            \left.\frac{\partial L}{\partial \dot x}\right|_{k} & = \lambda_{+,k}, & k = 1,\ldots,N-1 \\
            x_{+,k+1} & = x_{+,k} + \epsilon \dot x_{+,k+\frac12}, & k = 1,\ldots,N-1
        \end{aligned}
        \right.
        \\[1ex]
        \text{Lag.\ Mult.\ }\quad
        &
        \label{eq:discreteELfinallam}
        \lambda_{+,N} = \lambda_{+,N-1},
    \end{align}
\end{subequations}
where
\begin{equation}
    \begin{aligned}
        \left.\frac{\partial L}{\partial x}\right|_{k} & \equiv \left.\frac{\partial L}{\partial x}\right|_{\substack{\!\!\!\!\!\!\!\!x = x_{+,k} \\[3pt] \dot x = \dot x_{+,k+\frac12}}} \\[5pt]
        \left.\frac{\partial L}{\partial \dot x}\right|_{k} & \equiv \left.\frac{\partial L}{\partial \dot x}\right|_{\substack{\!\!\!\!\!\!\!\!x = x_{+,k} \\[3pt] \dot x = \dot x_{+,k+\frac12}}}.
    \end{aligned}
\end{equation}

The interpretation of the above is as follows.  The first line of \cref{eq:discreteELlamIC} sets the initial condition for $x_{+,k}$.  The following two lines of \cref{eq:discreteELlamIC} form a simultaneous set of equations to be solved for the initial conditions of $\dot x_{+,k+\frac12}$ and $\lambda_{+,k}$.  \cref{eq:discreteELx1} provides the first time step evolution in the position.  The following equations \cref{eq:discreteELxdEOM} evolve the degrees of freedom in time, with the last time step evolution giving the final position $x_{+,N}$.  The final equation \cref{eq:discreteELfinallam} absorbs the leftover (unmeasurable) Lagrange multiplier evolution.

As we have done before, we wish to examine the Hessian in order to ensure that we can appropriately apply the Method of Stationary Phase and also perform a non-trivial consistency check to find that the quantum expectation value is given in the $\hbar\rightarrow0$ limit by the classical path.

One may show after some effort that the Hessian is, to leading order in $\epsilon$, bipartite (and hence the signature is 0) with
\begin{align}
    \det\big( \Hess(S_{HR}) \big) = \epsilon^{2N}\prod_{i=0}^{N-1}\left( \left.\frac{\partial^2 L}{\partial \dot x_{i+\frac12}}\right|_{i} \right)^2 + \mathcal O(\epsilon^{2N-1}).
\end{align}
For our particular calculation, in which
\begin{align}
    L = \frac12m\dot x^2,
\end{align}
we therefore have that in the classical limit
\begin{align}
    \det\big( \Hess(S_{HR}) \big) = (m\epsilon)^{2N} + \mathcal O(\epsilon^{2N-1}).
\end{align}
One then finds that
\begin{align}
    \langle\hat x\rangle (t)
    & = \left( \frac{m\epsilon}{2\pi\hbar}\right)^N \left( \frac{1}{2\pi\hbar}\right)^{2N+2} \big[ (m\epsilon)^{2N}\big]^{-1/2}(2\pi\hbar)^{\frac12[2N+2+2n+2+2N]}x_{cl}(t) \nonumber\\
    & = x_{cl}(t),
\end{align}
where $x_{cl}(t)$ solves the usual Euler-Lagrange equations of motion subject to the initial values $x_{cl}(t_i) = x_0$ and $d x_{cl}/dt|_{t_i}=p_0/m$.

\subsection{Continuous Case: Configuration Space Path Integral}
\label{sec:continuousEL}
The continuum generalization of \cref{eq:discreteEL1,eq:discreteELaction} is exactly as one would expect:
\begin{align}
    \langle \hat x\rangle(t_f) & = \int\mathcal D[x_\pm(t),\dot x_{\pm}(t),\lambda_\pm(t)] e^{-\frac{x_-^2(t_i)}{8\sigma_x^2}}e^{-\frac{\sigma_x^2\lambda_-^2(t_i)}{2}} x_+(t_f) \nonumber\\
    & \qquad\qquad e^{\frac i\hbar S_{HR}[x_\pm(t),\dot x_{\pm}(t),\lambda_\pm(t)]}, \\
    S_{HR}[x_\pm(t),\dot x_{\pm}(t),\lambda_\pm(t)] & \equiv p_0 x_-(t_i) - \lambda_+(t_f)x_-(t_f) + \lambda_-(t_i)(x_+(t_i)-x_0) \nonumber\\
    & \quad + \int_{t_i}^{t_f} dt \big\{ 
        L[x_+(t) + \frac12x_-(t),\dot x_{+}(t) + \frac12 \dot x_{-}(t)] \nonumber\\
    & \qquad - L[x_+(t) - \frac12x_-(t),\dot x_{+}(t) - \frac12 \dot x_{-}(t)] \nonumber\\
    & \qquad +\lambda_{+}(t)\left( \frac{d}{dt}x_{-}(t) - \dot x_{-}(t)\right) \nonumber\\
    & \qquad + \lambda_{-}(t)\left( \frac{d}{dt}x_{+}(t) - \dot x_{+}(t) \right) \big\}.
\end{align}

The Method of Stationary Phase then gives in the $\hbar\rightarrow0$ limit $\delta S_{HR}=0$ in the now usual way.  In this case, by explicitly including the $\dot x_{\pm}(t)$ and $\lambda_{\pm}(t)$ variables in the path integral measure, we understand and have complete control over the variation of the action.  In particular, there is no confusion regarding whether the variations of $\dot x_{\pm}(t)$ are independent from the variations of $x_{\pm}(t)$: the $\dot x_{\pm}(t)$ variations \emph{are} independent of the $x_{\pm}(t)$ variations, with the connection between the $\dot x_{\pm}(t)$ variations and the $x_{\pm}(t)$ variations enforced by the Lagrange multipliers $\lambda_{\pm}(t)$ only \emph{after} the $\hbar\rightarrow0$ limit is taken.  Moreover, there is no ambiguity regarding the transposition rule: there is no need to invoke or consider the relationship between $\delta\dot x_\pm(t)$ and $\frac{d}{dt}[\delta x_\pm(t)]$ as these quantities are all independent of each other prior to taking the $\hbar\rightarrow0$ limit.  One might even hope that the methods developed here may resolve outstanding issues related to the classical mechanics and variational formulation of the mechanics of systems with general non-holonomic constraints \cite{Flannery2005_enigma_nonholonomic,Horowitz:2024eea,Bert:2025ffu}.

Again, in the usual way, the ``initial conditions'' and equations of motion for the minus coordinates are solved uniquely by $x_{-}(t)=\dot x_{-}(t)=\lambda_{-}(t)=0$.  Further, one finds, as expected, that the Lagrange multipliers can be thought of as the momenta of the system.  In particular, the $\hbar\rightarrow0$ limit fixes
\begin{align}
    \lambda_+ = \frac{\partial L}{\partial \dot x}.
\end{align}
We thus find that the $\hbar\rightarrow0$ limit gives the Euler-Lagrange equations as an \emph{initial value problem}.  
In particular, the initial conditions and equations of motion for the plus coordinates then become (dropping the pluses for notational clarity and to make a more ready connection to the usual classical mechanical formulation):
\begin{subequations}
    \begin{align}
        \text{\shortstack{Initial \\[5pt] Conditions}}\quad
        & \left\{
        \begin{aligned}
            x(t_i) & = x_0 \\
            \label{eq:xdotIC}
            \qquad\qquad\,\, \left.\frac{\partial L}{\partial \dot x}\right|_{t_i} & = p_0
        \end{aligned} \right. \\[1ex]
        \text{\shortstack{Equations \\[5pt] of Motion}}\quad
        & \left\{
        \begin{aligned}
            \dot x(t) & = \frac{d}{dt}\big[x(t)\big]\\
            \frac{d}{dt}\left( \frac{\partial L}{\partial \dot x} \right) - \frac{\partial L}{\partial x} & = 0.
        \end{aligned} \right.
    \end{align}
\end{subequations}
Note that, in the above, \cref{eq:xdotIC} is an implicit equation for the initial condition $\dot x(t_i)$.  

The consideration of the integration over the Gaussian fluctuations proceeds in an identical fashion as in \cref{sec:continuousHamilton}, with the functional integrals in the numerator and denominator exactly cancelling to yield
\begin{align}
    \langle\hat x\rangle(t_f) = x_{cl}(t_f).
\end{align}

\section{Connecting to Galley}
\label{sec:galley}
Now that we have completed our discussion of the derivation of Hamilton's Equations and the Euler-Lagrange Equations from the $\hbar\rightarrow0$ limit of the expectation value of the quantum position operator, we would like to compare the classical variational problem we derived to the classical variational problem posited by Galley \cite{Galley:2012hx}.  In \cite{Galley:2012hx}, there are no Lagrange multipliers and no injection of momentum $p_0$ from the initial quantum wavepacket.  In order for the contribution from the boundary terms (from integrating by parts the variation of his action) to be 0, Galley sets (in our language)
\begin{align}
    \label{eq:GalleyVariations}
    \delta x_-(t_f) & = 0 \nonumber\\
    \delta x_-(t_i) & = 0 \\
    \delta x_+(t_i) & = 0 \nonumber
\end{align}
and finds dynamically (i.e.\ that by setting the variation of the action to 0) that
\begin{align}
    \left.\frac{dx_-}{dt}\right|_{t_f} & = 0.
\end{align}
It is unclear how in \cite{Galley:2012hx} the initial momentum of the particle leads to an initial condition for the time derivative of the plus coordinate evaluated at the initial time.

There is a further, practical problem with \cref{eq:GalleyVariations}: setting $\delta x_-(t_f) = 0$, which leads dynamically to $\left.\frac{dx_-}{dt}\right|_{t_f} = 0$, and $\delta x_-(t_i) = 0$ then gives \emph{three} conditions for $x_-(t)$, which \emph{overdetermines} the $x_-(t)$ degree of freedom.  This overdetermination leads to jumps in the solutions at the final time slices of numerical implementations of the Schwinger-Keldysh-Galley variational action principle \cite{Rothkopf:2023ljz,Rothkopf:2024hxi}.

We saw in the $\hbar\rightarrow0$ limit of the Schwinger-Keldysh position space path integral in \cref{sec:continuouspositionPI} that 1) there is an injection of the initial momentum $p_0$ into the action to be varied and 2) there are only two path variations that are set to zero,
\begin{align}
    \begin{aligned}
        \delta x_-(t_f) & = 0 \\
        \delta x_+(t_i) & = 0.
    \end{aligned}
\end{align}
In particular, we do not find $\delta x_-(t_i)=0$.  It is precisely by allowing $x_-(t_i)$ to vary that one finds
\begin{align}
    \left. \frac{\partial L}{\partial\left(\frac{d}{dt}\big[x(t)\big]\right)} \right|_{t=t_i}  = p_0
\end{align}
as one of the two initial conditions for the classical equations of motion.  And by fixing only $\delta x_-(t_f)$ and $\delta x_+(t_i)$, we are able to find numerical solutions to the variational problem without jumps at the final time slices \cite{Rothkopf2026}.  

A minimal correction to the action principle proposed in \cite{Galley:2012hx} is to consider (in the language of \cite{Galley:2012hx})
\begin{align}
    \label{eq:correctedGalleyaction}
    S[\vec q_a] = \vec \pi_0\cdot \vec q_-(t_i) + \int_{t_i}^{t_f} dt \big[ L(\vec q_1,\dot\vec q_1) - L(\vec q_2,\dot\vec q_2) + K(\vec q_a,\dot\vec q_a)\big]
\end{align}
with
\begin{equation}
    \begin{aligned}
        \label{eq:correctedGalleyconditions}
        \vec q_+(t_i) & = \vec q_0 \\
        \vec q_-(t_f) & = \vec 0 \\
        \vec \eta_+(t_i) & = \vec 0 \\
        \vec \eta_-(t_f) & = \vec 0.
    \end{aligned}
\end{equation}
We have proven \cref{eq:correctedGalleyaction,eq:correctedGalleyconditions} to be the right correction to \cite{Galley:2012hx} when considering conservative forces.  A major advance of \cite{Galley:2012hx} was the incorporation of dissipative forces in a variational framework.  Presumably the above correction \cref{eq:correctedGalleyaction,eq:correctedGalleyconditions} will also work for non-conservative forces, even though we have not considered the classical limit of a quantum system with dissipation in this work.

\section{Method of Stationary Phase Considerations}
\label{sec:stationaryphase}
Let us now come back to the subtlety first noted below \cref{eq:expectationvaluepm1}, that the non-phase $g(x)$ in the integrand in our path integrals includes a dependence on $\hbar$, whereas the $g(x)$ in the usual formula for the application of the Method of Stationary Phase to a multidimensional integral \cref{eq:stationaryphase} is independent of the small parameter $a$.  One may worry that such a discrepancy may invalidate the Method of Stationary Phase result \cref{eq:stationaryphase} that we have extensively used throughout this work.  We will now explain why for our path integrals this discrepancy is unimportant.

Physically, the only dimensionful quantities in the problem are $\hbar,\,\sigma_x,\,\Delta t,\,x_0,$ and $p_0$; since the troublesome part of $g(x)$ in our path integrals is independent of all quantities except $\hbar$ and $\sigma_x$, it is not possible once all the measures are integrated over for the final result to be modified compared to the general formula \cref{eq:stationaryphase}.  One may back up this physical intuition by a careful consideration of the derivation of the Method of Stationary Phase from harmonic analysis; see, e.g., \cite{Wolff2003}.  Such a consideration shows that the steps that lead to the leading order in $a$ result \cref{eq:stationaryphase} from harmonic analysis are unmodified in the case that $g(x)$ depends on $a$.  

Although the naive leading order result is unmodified when $g(x)$ depends on $a$, one may still be concerned that the power counting is modified.  In our case, one may worry that either the leading order result is exponentially suppressed in $\hbar$ or that the higher order corrections are less suppressed.  Let us first consider whether the leading order contribution has a modified power counting.  We saw that the requirement of a stationary phase always led to ``initial'' conditions at the final time and equations of motion whose unique solution is zero for all the minus coordinates (including the Lagrange multipliers) for all time.  The offending term in our path integrals' $g(x)$ is a Gaussian of a minus coordinate of dimensions of momentum (either a Lagrange multiplier $\lambda_-$ or a momentum $p_-$) divided by $\hbar^2$ (and multiplied by $\sigma_x^2$, thus making the dimensions work out).  Thus when that minus momentum is replaced by 0, the Gaussian becomes equal to 1 independent of $\hbar$; thus the leading order contribution as given by \cref{eq:stationaryphase} is both unmodified and has unmodified power counting for our cases.  Let us now consider whether the higher order corrections to the leading order result of the Method of Stationary Phase have a modified power counting.  Intuitively, one expects that the presence of a Gaussian whose argument scales like $1/\hbar^2$ would lead to further suppression at higher orders.  Following again the steps of \cite{Wolff2003}, one can show that, indeed, higher order terms are suppressed by an additional power of $\hbar$ for our particular integrand.  (To be clear, in this case one can show that once $\hbar$ is scaled by a dimensionless number $a$, then the higher order terms are suppressed by an additional power of $a$.)

\section{Conclusions}
In this work we derived the classical limit of the Schwinger-Keldysh formalism for the expectation value of the non-relativistic quantum position operator in one dimension subject to a conservative force.  In so doing, we found the variational principle for initial value problems in classical point particle mechanics.  In particular, the variation of what we call Hamilton's Revised Action,
\begin{align}
    \delta S_{HR}|_{\substack{x_{+,cl}(t) \\ x_{-,cl}(t)}} = 0,
\end{align}
where
\begin{align}
    S_{HR}[x_\pm(t)] & \equiv p_0 x_-(t_i) + \int_{t_i}^{t_f} dt \, L\big( x_1(t), \frac{d}{dt}[x_1(t)] \big) - L\big( x_2(t), \frac{d}{dt}[x_2(t)] \big), \\
    x_+(t) & = \frac12\big( x_1(t) + x_2(t)\big) \\
    x_-(t) & = x_1(t) - x_2(t),
\end{align}
subject to
\begin{align}
    \begin{aligned}
        \delta x_-(t_f) & = 0 \\
        \delta x_+(t_i) & = 0
    \end{aligned}
\end{align}
yields the classical Hamilton's Equations and Euler-Lagrange equations as \emph{initial value problems}, where
\begin{align}
    \begin{aligned}
        x(t_i) & = x_0 \\
        p(t_i) & = p_0.
    \end{aligned}
\end{align}

We showed how in the above formulation, in which the action is written in terms of the explicit time derivative of the path, the variation of the action includes a \emph{functional chain rule} contribution from the Lagrangian's dependence on $\frac{d}{d}[x(t)]$; i.e.\ in this formulation the time derivative of the path is \emph{not} considered an independent variation from the variation of the path itself.

We then introduced Lagrange multipier functions to yield a configuration space Schwinger-Keldysh path integral.  The resulting Hamilton's Revised Action is
\begin{align}
    S_{HR}[x_\pm(t),\dot x_\pm(t),\lambda_\pm(t)] & \equiv p_0 x_-(t_i) - \lambda_+(t_f)x_-(t_f) + \lambda_-(t_i)\big( x_+(t_i) - x_0 \big) \nonumber \\
    & \quad+ \int_{t_i}^{t_f} dt \, L\big( x_1(t), \dot x_1(t) \big) - L\big( x_2(t), \dot x_2(t) \big) \nonumber\\
    & \qquad \lambda_+(t)\left( \left.\frac{dx_-}{dt}\right|_t - \dot x_-(t) \right) + \lambda_-(t)\left( \left.\frac{dx_+}{dt}\right|_t - \dot x_+(t) \right).
\end{align}
Then
\begin{align}
    \delta S_{HR}|_{\substack{x_{\pm,cl}(t) \\ \dot x_{\pm,cl}(t) \\ \lambda_{\pm,cl}(t)}} = 0,
\end{align}
where the variations $\delta x_\pm(t)$, $\delta \dot x_\pm(t)$, and $\delta \lambda_\pm(t)$ are \emph{all} independent and completely unconstrained, yield the classical Euler-Lagrange Equations as an initial value problem.  

A huge advantage of our approach is that determining classical mechanics from the $\hbar\rightarrow0$ limit of the quantum Schwinger-Keldysh path integral makes what is varied and how in the classical limit completely transparent and unambiguous.  Since non-holonomic constraints $g_k\big(q^i(t),\dot q^i(t),t)=0$ impose non-trivial relations between classical variations in (equivalently the quantum integration measure over) velocities and positions, we anticipate that this explicit control over the classical variations from the quantum path integral will be exceptionally useful and may allow for a fully general implementation of non-holonomic constraints in the path integral and in a variational action formulation of classical mechanics. 

In the above we were surprised to find that the classical limit always led to ``initial conditions'' at the final time slice for the minus coordinates, that the solution to the equations of motion for the minus coordinates propagated backwards in time, and that the unique solution to the equations of motion subject to the ``intial conditions'' set the minus coordinates identically to 0.

In all of the above, we started from a discrete time slicing, in which all quantities are under complete theoretical control.  The discrete time slicing results themselves are of interest in numerical implementations of our action principles and the resulting equations of motion.  We found, interestingly, that the inclusion of various Lagrange multipliers allowed for the discrete equations of motion to maintain their order of accuracy even at the temporal endpoints.  

The above procedures and techniques are very general.  For example, we have derived, but not shown here, identical results when considering the expectation value of the momentum operator.  In the appendix, we show how to treat $d$ generalized coordinates.  In the process of treating the generalized coordinates problem, we showed how to Trotterize terms in the Lagrangian that contain non-commuting operators without conjecturing mid- or post-point rules.  There should be no obstacle to further generalizing our results to field theories.  For field theories, non-holonomic constraints are common in the form of the oft-used Coulomb and Lorenz gauges and will thus form an interesting direction of future study.

\section{Acknowledgements}
WAH and AR thank J Nordstr\"om and M Sievert for valuable discussions.

This research was conducted in part by WAH while visiting the Okinawa Institute of Science and Technology (OIST) through the Theoretical Sciences Visiting Program (TSVP).  WAH thanks the National Research Foundation and the SA-CERN collaboration for their generous financial support during the course of this work.

AR gladly acknowledges support by Korea University through project K2510461 {\it Fully Dynamical Coordinate Maps for Space-Time Symmetry Preserving Lattice Field Theory} as well as project K2511131 and K2503291.

\section{Author Contributions}
WAH formulated the problem, performed all calculations, and wrote the first draft.  AR suggested the use of Lagrange multipliers to connect $\dot x(t)$ and $\frac{dx}{dt}|_t$, provided valuable discussions, and helped edit the manuscript.

\appendix
\section{Generalized Coordinates}
\label{sec:generalizedcoordinates}
One of the main advantages of the Hamiltonian and Lagrangian formulations in classical mechanics is the ability to use generalized coordinates $q^i(t)$ rather than Cartesian coordinates $x^i(t)$.  In \cref{sec:continuousHamilton,sec:continuousEL} we suggestively replaced $p^2/2m + V(x)$ by $H(p,q)$ and $m\dot x^2/2 - V(x)$ by $L(x,\dot x)$.  Under the assumption that the coordinates used were Cartesian, these replacements were justified.  However, we would like to carefully show that all our results hold for the usual types of general coordinates considered in classical mechanics.  In particular, we would like to consider our starting classical Lagrangian to be
\begin{align}
    \label{eq:generalL}
    L = \frac12mg_{ij}(\vec q)(\dot q^i \dot q^j) - V(\vec q), \qquad i = 1,\ldots,d.
\end{align}
One can think usefully of the coordinate-dependent quantity $g_{ij}$ as the metric of the coordinate space under consideration.  For Cartesian coordinates, $g_{ij}=\delta_{ij}$.

Given the Lagrangian \cref{eq:generalL}, we have that
\begin{align}
    p_i & = \frac{\partial L}{\partial \dot q^i} \nonumber\\
    & = m g_{ij} \dot q^j \\
    \Rightarrow \qquad \dot q^i & = \frac1m g^{ij} p_j,
\end{align}
where we define the inverse of $g_{ij}$ as
\begin{align}
    g^{ik}(\vec q) g_{kj}(\vec q) = \delta^i_j.
\end{align}

Then the classical Hamiltonian is given by
\begin{align}
    H(\vec p,\vec q) & = p_i\dot q^i - L(\vec q,\dot{\vec q}) \nonumber\\
    & = \frac{1}{2m}p_i g^{ij}(\vec q) p_j + V(\vec q).
\end{align}

We would now like to consider
\begin{align}
    \label{eq:generalavgx}
    \langle \hat x\rangle(t_f) = \langle \psi_i|e^{i\hat H(t_f-t_i)}\hat x e^{-i\hat H(t_f-t_i)}|\psi_i\rangle
\end{align}
given the generalized coordinates and Lagrangian given in \cref{eq:generalL}.  In order to compute \cref{eq:generalavgx}, we need to consider several generalizations of our previous work.  First, we should consider the decomposition of unity into complete sets of states needed for the Trotterization of \cref{eq:generalavgx}.  As noted above, we will usefully think of $g_{ij}$ as a metric, in which case we will choose to normalize our complete set of position states as
\begin{align}
    \label{eq:generaldecompositionofunityposition}
    \hat{\mathds{1}} & \equiv \int d^d q\sqrt{g(\vec q)}|\vec q\rangle\langle \vec q| \\
    \langle\vec q|\vec r\rangle & = \frac{1}{\sqrt{g(\vec q)}}\delta^{(d)}(\vec q- \vec r).
\end{align}
The decomposition of unity in momentum eigenstates will be
\begin{align}
    \hat{\mathds{1}} & = \int d^d p|\vec p\rangle\langle\vec p| \\
    \langle\vec p|\vec k\rangle & = \delta^{(d)}(\vec p - \vec k).
\end{align}
Notice how $g_{ij}$ does \emph{not} appear in the momentum space decomposition of unity or inner product; $g_{ij}$ \emph{cannot} appear in the momentum space inner product or decomposition of unity because $g_{ij}$ by assumption depends on the coordinates $q^i$, which cannot be incorporated into the momentum space inner product or decomposition of unity, which depend on the components of the momentum $p_i$ \emph{only}.  Note that we may consider coordinates of finite extent (as opposed to infinite extent), in which case the momenta will be discretized; in the case of discrete momenta, one could either replace the integrals with sums, or, more usefully (as we'll soon be considering the $\hbar\rightarrow0$ limit), consider the integrals to be in the Lebesgue rather than Riemann sense.

With the above conventions, the usual Dirac Quantization Condition
\begin{align}
    [\hat q^i,\hat p_j] = i\hbar\delta^i_j
\end{align}
implies that
\begin{align}
    \label{eq:generalpositionmomentumoverlap}
    \langle \vec q|\vec p\rangle = \frac{1}{(2\pi\hbar)^{d/2}}\frac{1}{\sqrt[4]{g(\vec q)}}e^{\frac i\hbar \vec q\cdot\vec p},
\end{align}
where $\vec q\cdot\vec p \equiv q^ip_i$.  The fourth root of the determinant of $g_{ij}$ looks strange in \cref{eq:generalpositionmomentumoverlap}, but is there to cancel the square root of the determinant of $g_{ij}$ in the decomposition of unity in the position eigenbasis, \cref{eq:generaldecompositionofunityposition}.

Next, we should consider the generalization of the initial wavepacket.  We will take the position-space representation of the initial wavepacket to be
\begin{align}
    \label{eq:curvedspacewavepacket}
    \langle \vec q|\psi_0\rangle = \frac{1}{\sqrt[4]{g(\vec q)(2\pi)^d\sigma^{-2}}}e^{-\frac14(q^i-q_0^i)\sigma_{ij}^2(q^j-q_0^j)+\frac i\hbar q^ip_{0,i}+\mathcal O(R\sigma^3)},
\end{align}
where $g(\vec q)\equiv\det g_{ij}(\vec q)$, $\sigma^{-2}\equiv\det[(\sigma^2)^{ij}]$, $(\sigma^2)^{ik}\sigma^2_{kj}=\delta^i_j$, and $R$ denotes curvature terms associated with the $g_{ij}$.  This generalized wavepacket satisfies our usual expectations:
\begin{align}
    \langle \psi_0|\psi_0\rangle & = 1 \nonumber\\ 
    \langle \psi_0|\hat{\vec q}|\psi_0\rangle & = \vec q_0 \nonumber\\ 
    \langle \psi_0|(\hat q^i - q_0^i)(\hat q^j - q_0^j)|\psi_0\rangle & = (\sigma^2)^{ij} \\ 
    \langle \psi_0|\hat p_i|\psi_0\rangle & = p_i \nonumber\\
    \langle \psi_0|(\hat p_i - p_{0,i})(\hat p_j - p_{0,j})|\psi_0\rangle & = \frac{\hbar^2}{4}\sigma^2_{ij}.\nonumber
\end{align}

Finally, we should consider the operator ordering ambiguity of the kinetic term in the Hamiltonian.  Let us elevate the Hamiltonian to the operator
\begin{align}
    \label{eq:generalHop}
    \hat H = \frac{1}{2m}\hat p_i g^{ij}(\hat{\vec q}) \hat p_j + V(\hat{\vec q}).
\end{align}
\cref{eq:generalHop} is the unique elevation of the classical kinetic term to an operator that ensures Hermicity and coordinate invariance, up to higher order corrections in curvature $R$ \cite{DeWitt:1957at}.  The higher order corrections in $R$ are $\mathcal O(\hbar^2)$ and are thus irrelevant for our classical $\hbar\rightarrow0$ considerations.

We now Trotterize \cref{eq:generalavgx} using the above:
\begin{align}
    \langle \hat{\vec q}\rangle(t_f)
    & = \lim_{N\rightarrow\infinity}\langle\psi_0|\underbrace{e^{i\hat H\epsilon}\cdots e^{i\hat H\epsilon}}_{N\text{ times}}\hat{\vec q}\underbrace{e^{-i\hat H\epsilon}\cdots e^{-i\hat H\epsilon}}_{N\text{ times}}|\psi_0\rangle, \qquad \epsilon\equiv \frac{t_f-t_i}{N}.
\end{align}
Then insert $2(N+1)$ decompositions of unity,
\begin{align}
    \langle\hat{\vec{q}}\rangle(t)
    & = \lim_{N\rightarrow\infinity}\prod_{n_1 = 0}^{N} \int d^d q_{1,n_1}\sqrt{g(\vec q_{1,n_1})} \prod_{n_2 = 0}^{N} \int d^d q_{2,n_2}\sqrt{g(\vec q_{2,n_2})} \nonumber\\
    &\qquad \langle\psi_0|\vec q_{2,0}\rangle\langle\vec q_{2,0} | e^{i\hat H\epsilon}|\vec q_{2,1}\rangle\ldots\langle \vec q_{2,N}|\hat{\vec q}|\vec q_{1,N}\rangle\ldots\langle \vec q_{1,1}|e^{-i\hat H\epsilon}|\vec q_{1,0}\rangle\langle \vec q_{1,0}|\psi_0\rangle.
\end{align}
We now need to evaluate
\begin{align}
    \langle \vec q_{1,k+1}| e^{-i\hat H\epsilon}|\vec q_{1,k} \rangle
    & = \langle \vec q_{1,k+1}| e^{-i\epsilon\big[ \frac12V(\hat{\vec q}) + \frac{1}{2m}\hat p_i g^{ij}(\hat{\vec q}) \hat p_j + \frac12V(\hat{\vec q}) \big]}|\vec q_{1,k} \rangle \nonumber\\
    & = \langle \vec q_{1,k+1}| e^{-i\epsilon \frac12V(\hat{\vec q})}e^{-i\epsilon\frac{1}{2m}\hat p_i g^{ij}(\hat{\vec q}) \hat p_j}e^{-i\epsilon\frac12V(\hat{\vec q})}|\vec q_{1,k} \rangle +\mathcal O(\epsilon^3).
\end{align}
In the first line, we have chosen to consider the Strang-split Hamiltonian, with the potential split equally before and after the kinetic term.  Choosing to Strang-split allows a more symmetric treatment of the matrix elements of the time evolution operator.  In principle, Strang-splitting also has the advantage of being second order accurate in $\epsilon$. 

The position eigenstates trivially pass through the potential terms.  We are thus left to evaluate the position space matrix element of the kinetic term.  Classic treatments of path integrals either assert a midpoint rule \cite{FeynmanHibbs1965,Marinov:1980vn}, claim that the midpoint rule is the ``simplest'' \cite{Schulman:1981vu}, or assert a new quantum equivalence principle and utilize a postpoint rule \cite{Kleinert2009}.  In a way that we have not seen elsewhere in the literature, we will rather carefully determine the discretized time evolution operator matrix element by expanding in $\epsilon$ and then inserting complete sets of \emph{both} position and momentum eigenstates as necessary.  We will see that this procedure does \emph{not} lead to the usual midpoint rule as applied to the kinetic term.  (Note again that, at least to quadratic order in $\hbar$, the operator ordering of the kinetic term is uniquely fixed by unitarity, which is to say the Hermiticity of the time evolution operator.)  Specifically, we need to consider
\begin{multline}
    \langle \vec q_{1,k+1}| \hat p_i g^{ij}(\hat{\vec q}) \hat p_j|\vec q_{1,k} \rangle
    = \int d^dq_{1,k+\frac12} \sqrt{g(\vec q_{1,k+\frac12})} \int d^dp_{1,k+\frac14} \int d^dp_{1,k+\frac34} \\
    \langle \vec q_{1,k+1}|\hat p_i|\vec p_{1,k+\frac34}\rangle\langle \vec p_{1,k+\frac34}|\vec q_{1,k+\frac12}\rangle\langle \vec q_{1,k+\frac12}|g^{ij}(\hat{\vec q})|\vec p_{1,k+\frac14}\rangle\langle \vec p_{1,k+\frac14}|\hat p_j|\vec q_{1,k} \rangle,
\end{multline}
which leads, after some evaluations and rearrangements, to
\begin{align}
    \label{eq:generaltimeevolutionmatrixelement}
    \langle \vec q_{1,k+1}| e^{-i\hat H\epsilon}|\vec q_{1,k} \rangle
    & = \frac{1}{\sqrt[4]{g(q_{1,k+1})}}e^{-i\epsilon\frac12V(\vec q_{1,k+1})} \nonumber\\
    & \quad \int d^dq_{1,k+\frac12} \int \frac{d^dp_{1,k+\frac14}}{(2\pi\hbar)^d}\frac{d^dp_{1,k+\frac34}}{(2\pi\hbar)^d}  \nonumber\\
    & \qquad e^{\frac i\hbar \vec p_{1,k+\frac34}\cdot(\vec q_{1,k+1} - \vec q_{1,k+\frac12})}e^{\frac i\hbar \vec p_{1,k+\frac14}\cdot(\vec q_{1,k+\frac12} - \vec q_{1,k})} \nonumber\\
    & \qquad e^{-\frac i\hbar \epsilon\frac{1}{2m}p_{i;1,k+\frac34}g^{ij}(\vec q_{1,k+\frac12})p_{j;1,k+\frac14}} \nonumber\\
    & \quad \frac{1}{\sqrt[4]{g(q_{1,k})}}e^{-i\epsilon\frac12V(\vec q_{1,k})} + \mathcal O(\epsilon^2).
\end{align}
Note first that in \cref{eq:generaltimeevolutionmatrixelement} the indications of specific intermediate time steps---the position evaluated at a half time step and the momenta evaluated at quarter time steps---do not have any justification and are there mainly as a bookkeeping device; the only physical time steps are the integer ones, which, in our choice of Strang splitting, are associated with the evaluations of the potentials.  Note further that by only expanding the exponential of the kinetic term to first order in $\epsilon$, we have reduced the overall order of accuracy of the time evolution operator matrix element to first order in $\epsilon$.  

One then finds that
\begin{align}
    \label{eq:general12phasespacepathintegral}
    \langle\vec q\rangle(t)
    & = \lim_{N\rightarrow\infinity}\int d^d q_{1,N} d^dq_{2,N} \nonumber\\
    & \quad \prod_{n=0}^{N-1} \int d^d q_{1,n} \int d^d q_{2,n} d^d q_{1,n+\frac12} \frac{d^d p_{1,n+\frac14}}{(2\pi\hbar)^d}\frac{d^d p_{1,n+\frac34}}{(2\pi\hbar)^d} d^d q_{2,n+\frac12} \frac{d^d p_{2,n+\frac14}}{(2\pi\hbar)^d}\frac{d^d p_{2,n+\frac34}}{(2\pi\hbar)^d} \nonumber\\
    & \qquad \frac{1}{\sqrt[4]{(2\pi)^d\sigma^{-2}}}e^{-\frac14(q_{2,0}^i - q_0^i)\sigma_{ij}^2(q_{2,0}^j - q_0^j) - \frac{i}{\hbar}p_{i;0}q_{2,0}^i} \nonumber\\
    & \qquad \frac{1}{\sqrt[4]{(2\pi)^d\sigma^{-2}}}e^{-\frac14(q_{1,0}^i - q_0^i)\sigma_{ij}^2(q_{1,0}^j - q_0^j) + \frac{i}{\hbar}p_{i;0}q_{1,0}^i} \nonumber\\
    & \qquad \exp\left[ \frac i\hbar \epsilon \sum_{m_2=0}^{N-1} \frac{1}{2m}p_{i;2,m_2+\frac34}g^{ij}(\vec q_{2,m_2+\frac12})p_{j;2,m_2+\frac14} \right. \nonumber\\
    & \qquad\quad - \frac12p_{i;2,m_2+\frac34}\frac{q_{2,m_2+1}^i - q_{2,m_2+\frac12}^i}{\epsilon/2} - \frac12p_{i;2,m_2+\frac14}\frac{q_{2,m_2+\frac12}^i - q_{2,m_2}^i}{\epsilon/2} \nonumber\\
    & \qquad\quad\left.+V(\vec q_{2,m_2})\big(1-\frac12\delta_{m_2,0}\big) \right]e^{\frac i\hbar\epsilon\frac12V(\vec q_{2,N})} \nonumber\\
    & \qquad \exp\left[ -\frac i\hbar \epsilon \sum_{m_1=0}^{N-1} \frac{1}{2m}p_{i;1,m_1+\frac34}g^{ij}(\vec q_{1,m_1+\frac12})p_{j;1,m_1+\frac14} \right. \nonumber\\
    & \qquad\quad - \frac12p_{i;1,m_1+\frac34}\frac{q_{1,m_1+1}^i - q_{1,m_1+\frac12}^i}{\epsilon/2} - \frac12p_{i;1,m_1+\frac14}\frac{q_{1,m_1+\frac12}^i - q_{1,m_1}^i}{\epsilon/2} \nonumber\\
    & \qquad\quad\left.+V(\vec q_{1,m_1})\big(1-\frac12\delta_{m_1,0}\big) \right]e^{-\frac i\hbar\epsilon\frac12V(\vec q_{1,N})} \nonumber\\
    & \qquad \vec q_{1,N}\delta^{(d)}(\vec q_{2,N} - \vec q_{1,N}).
\end{align}
After changing $+/-$ coordinates and introducing
\begin{align}
    \delta^{(d)}(\vec q_{-,N}) & = \int \frac{d^d \lambda_+}{(2\pi\hbar)^d} e^{-i\lambda_{i,+}q_{-,N}^i} \nonumber\\
    \exp\big[ -\frac12(q_{+,0}^i-q_0^i)\sigma_{ij}^2(q_{+,0}^j-q_0^j)\big] & = \frac{\sqrt{\sigma^{-2}}}{(2\pi)^{d/2}} \int d^d\lambda_- e^{-\frac12\lambda_{i,-}(\sigma^{-2})^{ij}\lambda_{j,-} + i\lambda_{i,-}(q_{+,0}^i - q_0^i)},
\end{align}
one finds that
\begin{align}
    \label{eq:generalHRpathintegral}
    \langle\vec q\rangle(t)
    & = \lim_{N\rightarrow\infinity}\int d^d q_{+,N} d^d q_{-,N} \frac{d^d \lambda_+}{(2\pi\hbar)^d}\frac{d^d \lambda_-}{(2\pi\hbar)^d} \, \vec q_{+,N} \nonumber\\
    & \quad \prod_{n=0}^{N-1} \int d^d q_{+,n} d^d q_{-,n} d^d q_{+,n+\frac12} d^d q_{-,n+\frac12} \frac{d^d p_{+,n+\frac14}}{(2\pi\hbar)^d}\frac{d^d p_{-,n+\frac34}}{(2\pi\hbar)^d}\frac{d^d p_{+,n+\frac14}}{(2\pi\hbar)^d}\frac{d^d p_{-,n+\frac34}}{(2\pi\hbar)^d} \nonumber\\
    & \qquad \exp\left[ -\frac18q_{-,0}^i\sigma_{ij}^2q_{-,0}^j-\frac12\lambda_{i,-}(\sigma^{-2})^{ij}\lambda_{j,-}\right]e^{\frac i\hbar S_{HR}[\vec q_{\pm,k}\vec q_{\pm,k+\frac12},\vec p_{\pm,k+\frac14},\vec p_{\pm,k+\frac34},\lambda_\pm]},
\end{align}
where
\begin{align}
    \label{eq:generalHRaction}
    S_{HR}&[\vec q_{\pm,k},\vec q_{\pm,k+\frac12},\vec p_{\pm,k+\frac14},\vec p_{\pm,k+\frac34},\lambda_\pm] \nonumber\\
    & \equiv \lambda_{i;-}(q_{+,0}^i-q_0^i) - \lambda_{i;+}q_{-,N}^i + p_{i;0}q_{-,0}^i + \frac12\epsilon\left[ V(\vec q_{+,N} - \frac12\vec q_{-,N}) - V(\vec q_{+,N} + \frac12\vec q_{-,N})\right] \nonumber\\
    & \quad +\epsilon\sum_{k=0}^{N-1} \frac{1}{2m}(p_{i;+,k+\frac34}p_{j;+,k+\frac14} + \frac14 p_{i;-,k+\frac34}p_{j;-,k+\frac14}) \nonumber\\
    & \qquad\qquad\qquad\qquad\qquad\qquad \times\left[ g^{ij}(\vec q_{+,k+\frac12}-\frac12\vec q_{-,k+\frac12}) - g^{ij}(\vec q_{+,k+\frac12}+\frac12\vec q_{-,k+\frac12})\right] \nonumber\\
    & \qquad-\frac{1}{4m}(p_{i;-,k+\frac34}p_{j;+,k+\frac14} + p_{i;+,k+\frac34}p_{j;-,k+\frac14}) \nonumber\\
    & \qquad\qquad\qquad\qquad\qquad\qquad \times\left[ g^{ij}(\vec q_{+,k+\frac12}-\frac12\vec q_{-,k+\frac12}) + g^{ij}(\vec q_{+,k+\frac12}+\frac12\vec q_{-,k+\frac12})\right] \nonumber\\
    & \qquad + \frac12 p_{i;+,k+\frac34}\frac{q_{-,k+1}^i - q_{-,k+\frac12}^i}{\epsilon/2} + \frac12 p_{i;+,k+\frac14}\frac{q_{-,k+\frac12}^i - q_{-,k}^i}{\epsilon/2} \nonumber\\
    & \qquad + \frac12 p_{i;-,k+\frac34}\frac{q_{+,k+1}^i - q_{+,k+\frac12}^i}{\epsilon/2} + \frac12 p_{i;-,k+\frac14}\frac{q_{+,k+\frac12}^i - q_{+,k}^i}{\epsilon/2} \nonumber\\
    & \qquad + \left[ V(\vec q_{+,k}-\frac12\vec q_{-,k}) - V(\vec q_{+,k}+\frac12\vec q_{-,k})\right](1-\frac12\delta_{k,0}).
\end{align}
Notice at this stage that $p_{i;\pm,k+\frac14}$ and $p_{i;\pm,k+\frac34}$ are completely independent degrees of freedom, which will each have its own equation of motion in the classical limit.  For $g_{ij}$ independent of coordinates, e.g.\ for $g_{ij}=\delta_{ij}$ Cartesian coordinates, then 1) the first line of the sum becomes identically 0 and 2) one can integrate out the $q_{\pm,k+\frac12}^i$, setting $p_{i;\pm,k+\frac14} = p_{i;\pm,k+\frac34}\equiv p_{i;\pm,k+\frac12}$; one then recovers the usual (multidimensional) Strang-split phase space path integral generalization of \cref{eq:expectationvaluepm0}. 

If one considers the variation of the action \cref{eq:generalHRaction}, setting to zero all terms proportional to the minus coordinates (assuming that, as was seen in all previous cases, the classical limit leads to initial conditions and equations of motion that lead to the unique solution of all minus variables being identically zero), then one finds the following for the plus coordinates (where we suppress the plus coordinates for notational clarity):
\begin{subequations}
    \label{eq:generaldiscreteHamiltonsEqs}
    \begin{align}
    \text{\shortstack{Initial \\[5pt] Conditions}}\quad
    &\left\{
    \begin{aligned}
        \label{eq:generalq0IC}
        \quad q_0^i & = q_0^i \\
        p_{i;\frac14} & = p_{i;0} -\frac\epsilon2\left. \frac{\partial V}{\partial q^i} \right|_0 \\
        q^i_{\frac12} & = q_0^i + \frac{\epsilon}{2m}p_{j;\frac34}\left.g^{ij}\right|_{\frac12} \\
        p_{i;\frac34} & = p_{i;\frac14} - \frac{\epsilon}{2m}p_{\ell;\frac34}p_{m;\frac14}\left. \frac{\partial g^{\ell m}}{\partial q^i} \right|_{\frac12}
    \end{aligned}
    \right. \\[1ex]
    \text{First Evolution} \quad
    & \quad
        \label{eq:generalqfirststep}
        q^i_{1} = q_{\frac12}^i + \frac{\epsilon}{2m}p_{j;\frac14}\left.g^{ij}\right|_{\frac12} \\[1ex]
    \text{\shortstack{Equations \\[5pt] of Motion}}\quad
    &\left\{
    \begin{aligned}
        \label{eq:generalp14kstep}
        \quad p_{i;k+\frac14} & = p_{i;k-1+\frac34} - \epsilon\left. \frac{\partial V}{\partial q^i} \right|_k, & k = 1,\dots,N-1 \\
        q^i_{k+\frac12} & = q_k^i + \frac{\epsilon}{2m}p_{j;k+\frac34}\left.g^{ij}\right|_{k+\frac12}, & k = 1,\dots,N-1 \\
        p_{i;k+\frac34} & = p_{i;k+\frac14} - \frac{\epsilon}{2m}p_{\ell;k+\frac34}p_{m;k+\frac14}\left. \frac{\partial g^{\ell m}}{\partial q^i} \right|_{k+\frac12}, & k = 1,\dots,N-1 \\
        q^i_{k+1} & = q_{k+\frac12}^i + \frac{\epsilon}{2m}p_{j;k+\frac34}\left.g^{ij}\right|_{k+\frac12}, & k = 1,\dots,N-1.
    \end{aligned}
    \right.
    \end{align}
\end{subequations}
In \cref{eq:generaldiscreteHamiltonsEqs}, \cref{eq:generalq0IC} are the initial conditions.  (The notation is not ideal in the first line: the left hand side is the value of the $i^\mathrm{th}$ component of the coordinate $\vec q$ at the $t_0$ timestep; the right hand side is the initial value of the position vector of the particle as determined by the experimental setup.)  As written, the last two lines of \cref{eq:generalq0IC} form a pair of implicit, coupled equations that must be solved simultaneously for $q_{\frac12}^i$ and $p_{i;\frac34}$.  \cref{eq:generalqfirststep} then gives the first time step evolution of the generalized coordinates.  \cref{eq:generalp14kstep} then give the subsequent time step evolution of the generalized coordinates and momenta.  The middle two lines of \cref{eq:generalp14kstep} form a pair of implicit, coupled equations that must be solved simultaneously for $q_{k+\frac12}^i$ and $p_{i;k+\frac34}$ at each time step. Finally, the last line of \cref{eq:generalp14kstep} gives the subsequent time step evolution for the generalized coordinates.  

Note that to $\mathcal O(\epsilon^2)$ accuracy, one may replace the $p_{j;k+\frac34}$ with $p_{j;k+\frac14}$ and $\left. g^{ij}\right|_{k+\frac12}$ with $\left. g^{ij}\right|_k$ in the second line of \cref{eq:generalp14kstep} and replace $p_{\ell;k+\frac34}$ with $p_{\ell;k+\frac14}$ and $\left.\frac{\partial g^{\ell m}}{\partial q^i}\right|_{k+\frac12}$ with $\left.\frac{\partial g^{\ell m}}{\partial q^i}\right|_k$ in the third line of \cref{eq:generalp14kstep}, and similarly for the coupled equations in the initial conditions \cref{eq:generalq0IC}.  With these replacements, \cref{eq:generalq0IC} and \cref{eq:generalp14kstep} all become explicit and uncoupled.  Then, in the $\epsilon\rightarrow0$ limit, one can see that \cref{eq:generaldiscreteHamiltonsEqs} yield the usual Hamilton's Equations in generalized coordinates as an initial value problem:
\begin{equation}
    \label{eq:generalcontinuousHamiltonsEqs}
    \begin{aligned}
        q_0^i & = q_0^i \\
        p_{i;0} & = p_{i;0} \\
        \frac{d q^i}{dt} & = \frac{\partial H}{\partial p^i} \\
        \frac{d p_i}{dt} & = -\frac{\partial H}{\partial q^i}.        
    \end{aligned}
\end{equation}

One may make the connection to the continuum case more explicit by considering the continuum version of the phase space path integral and action, \cref{eq:generalHRpathintegral,eq:generalHRaction}:
\begin{align}
    \langle\vec q\rangle(t_f) & = \int \mathcal D[\vec q_\pm(t),\vec p_{\pm}(t),\vec \lambda_\pm(t)] e^{-\frac18\vec q_-(t_i)\cdot\overleftrightarrow{\sigma}^2\cdot\vec q_-(t_i) - \frac12\vec \lambda_-(t_i)\cdot\overleftrightarrow{\sigma}^{-2}\cdot\vec \lambda_-(t_i)} e^{iS_{HR}[\vec q_\pm(t),\vec p_{\pm}(t),\vec \lambda_\pm(t)]},
\end{align}
where
\begin{align}
    S_{HR}&[\vec q_\pm(t),\vec p_{\pm}(t),\vec \lambda_\pm(t)] \nonumber\\
    & = \vec\lambda_-(t_i)\cdot(\vec q_+(t_i) - \vec q_0) - \vec\lambda_+(t_f)\cdot\vec q_-(t_f) + \vec p_0\cdot\vec q_-(t_i) \nonumber\\
    & \quad + \int_{t_i}^{t_f} dt \left[ \frac{1}{2m}(p_{i;+}p_{j;+} + \frac14 p_{i;-}p_{j;-})\left( g^{ij}(\vec q_+ - \frac12\vec q_-) - g^{ij}(\vec q_+ + \frac12\vec q_-)\right) \right. \nonumber\\
    & \qquad -\frac{1}{4m}(p_{i;-}p_{j;+} + p_{i;+}p_{j;-})\left( g^{ij}(\vec q_+ - \frac12\vec q_-) + g^{ij}(\vec q_+ + \frac12\vec q_-)\right) \nonumber\\
    & \qquad + \vec p_{+} \cdot \frac{d\vec q_-}{dt} + \vec p_{-}\cdot \frac{d\vec q_+}{dt} 
    \left. + V(\vec q_+ - \frac12\vec q_-) - V(\vec q_+ + \frac12\vec q_-)\right].
\end{align}
Then, considering only the variations in the minus coordinates (assuming that the variations in the plus coordinates will yield, as usual, to all the minus coordinates dynamically set to zero), one finds the usual Hamilton's Equations in generalized coordinates as an initial value problem, exactly as seen in the continuum limit of the discrete case above \cref{eq:generalcontinuousHamiltonsEqs}.

In order to derive the configuration space path integral formulation, and thus the Euler-Lagrange equations in the $\hbar\rightarrow0$ limit, we need to integrate out the momenta in \cref{eq:generalHRpathintegral}.  Trying to integrate out the momenta in the $\pm$ coordinate system is quite complicated due to the mixing of the different momenta through the $g^{ij}$.  Instead, it is much simpler to return to the phase space path integral in the original $1/2$ coordinate system, \cref{eq:general12phasespacepathintegral}.  One may then, e.g., readily integrate out the $p_{i;1,k+\frac34}$ momenta, leading to delta functions
\begin{align}
    \delta^{(d)}\left( \frac{\epsilon}{2m}p_{j;1,k+\frac14}g^{ij}(\vec q_{1,k+\frac12})-(q^i_{1,k+1} - q^i_{1,k+\frac12}) \right),
\end{align}
which may themselves be used to collapse the integrals over the $p_{i;1,k+\frac14}$ momenta.  Repeating the procedure for the $p_{i;2,k+\frac34}$ and $p_{i;2,k+\frac14}$ momenta, one finds
\begin{align}
    \langle\vec q\rangle(t)
    & = \lim_{N\rightarrow\infinity}\int d^d q_{1,N} d^d q_{2,N} \, \vec q_{1,N} \delta^{(d)}(\vec q_{1,N} - \vec q_{2,N}) \nonumber\\
    & \quad \prod_{n=0}^{N-1} \int d^d q_{1,n} \int d^d q_{2,n} d^d q_{1,n+\frac12} \frac{1}{(2\pi\hbar)^2}d^d q_{2,n+\frac12}\frac{1}{(2\pi\hbar)^2} \nonumber\\
    & \quad \left|\det\left[ \frac{\epsilon}{2m}g^{ij}(\vec q_{1,n+\frac12})\right]\right|^{-1} \left|\det\left[ \frac{\epsilon}{2m}g^{ij}(\vec q_{2,n+\frac12})\right]\right|^{-1} \nonumber\\
    & \qquad \frac{1}{\sqrt[4]{(2\pi)^d\sigma^{-2}}}e^{-\frac14(q_{2,0}^i - q_0^i)\sigma_{ij}^2(q_{2,0}^j - q_0^j) - \frac{i}{\hbar}p_{i;0}q_{2,0}^i} \nonumber\\
    & \qquad \frac{1}{\sqrt[4]{(2\pi)^d\sigma^{-2}}}e^{-\frac14(q_{1,0}^i - q_0^i)\sigma_{ij}^2(q_{1,0}^j - q_0^j) + \frac{i}{\hbar}p_{i;0}q_{1,0}^i} \nonumber\\
    & \qquad \exp\left[ \frac{i}{\hbar}\epsilon\sum_{m_2=0}^{N-1} \left( -\frac{m}{2}g_{ij}(\vec q_{2,m_2+\frac12})\frac{q_{2,m_2+1}^i - q_{2,m_2+\frac12}^i}{(\epsilon/2)}\frac{q_{2,m_2+\frac12}^i - q_{2,m_2}^i}{(\epsilon/2)} \right.\right. \nonumber\\
    & \qquad\qquad\qquad\qquad\qquad\qquad\qquad\qquad\qquad\qquad\qquad\qquad \left.\left. + V(\vec q_{2,m_2})(1-\frac12\delta_{m_2,0}) \right) \right] \nonumber\\
    & \qquad e^{\frac i\hbar\epsilon\frac12V(\vec q_{2,N})} \nonumber\\
    & \qquad \exp\left[ -\frac{i}{\hbar}\epsilon\sum_{m_1=0}^{N-1} \left( -\frac{m}{2}g_{ij}(\vec q_{1,m_1+\frac12})\frac{q_{1,m_1+1}^i - q_{1,m_1+\frac12}^i}{(\epsilon/2)}\frac{q_{1,m_1+\frac12}^i - q_{1,m_1}^i}{(\epsilon/2)} \right.\right. \nonumber\\
    & \qquad\qquad\qquad\qquad\qquad\qquad\qquad\qquad\qquad\qquad\qquad\qquad \left.\left. + V(\vec q_{1,m_1})(1-\frac12\delta_{m_1,0}) \right) \right] \nonumber\\
    & \qquad e^{-\frac i\hbar\epsilon\frac12V(\vec q_{1,N})}.
\end{align}
We may now change coordinates back to $\pm$, insert the Lagrange multipliers to convert the delta function over the final minus coordinates and the initial Gaussian smearing over the initial plus position, and insert the Lagrange multipliers to connect the time derivatives of the coordinates to the dotted coordinates.  Then the Hessian from the Method of Stationary Phase will yield precisely the correct determinants to cancel the determinants introduced by integrating out the $p_{i;1/2,k+\frac34}$ momenta and one will obtain the correct Euler-Lagrange equations as an initial value problem in generalized coordinates.  The continuum limit can be readily considered in a similar way.

\bibliography{Crefs}

@book{Arnold1989,
  author    = {V. I. Arnold},
  title     = {Mathematical Methods of Classical Mechanics},
  edition   = {2},
  series    = {Graduate Texts in Mathematics},
  volume    = {60},
  publisher = {Springer},
  address   = {New York},
  year      = {1989}
}

@article{Landsman:1986uw,
    author = "Landsman, N. P. and van Weert, C. G.",
    title = "{Real and Imaginary Time Field Theory at Finite Temperature and Density}",
    reportNumber = "ITFA-86-12",
    doi = "10.1016/0370-1573(87)90121-9",
    journal = "Phys. Rept.",
    volume = "145",
    pages = "141",
    year = "1987"
}

@article{Caldeira:1982iu,
    author = "Caldeira, A. O. and Leggett, A. J.",
    title = "{Path integral approach to quantum Brownian motion}",
    doi = "10.1016/0378-4371(83)90013-4",
    journal = "Physica A",
    volume = "121",
    pages = "587--616",
    year = "1983"
}

@article{Feynman:1963fq,
    author = "Feynman, R. P. and Vernon, Jr., F. L.",
    editor = "Brown, L. M.",
    title = "{The Theory of a general quantum system interacting with a linear dissipative system}",
    doi = "10.1016/0003-4916(63)90068-X",
    journal = "Annals Phys.",
    volume = "24",
    pages = "118--173",
    year = "1963"
}

@book{MorseFeshbach1953,
  author    = {P. M. Morse and H. Feshbach},
  title     = {Methods of Theoretical Physics},
  publisher = {McGraw-Hill},
  address   = {New York},
  year      = {1953},
  note      = {Section 3.2}
}

@book{Goldstein2002,
  author    = {Herbert Goldstein and Charles Poole and John Safko},
  title     = {Classical Mechanics},
  edition   = {3},
  publisher = {Addison-Wesley},
  address   = {San Francisco},
  year      = {2002}
}

@article{Noether1918,
  author  = {Emmy Noether},
  title   = {Invariante Variationsprobleme},
  journal = {Nachrichten von der Gesellschaft der Wissenschaften zu Göttingen, Mathematisch-Physikalische Klasse},
  pages   = {235--257},
  year    = {1918}
}

@article{RothkopfHorowitz2024APS,
  author  = {Alexander Rothkopf and W. A. Horowitz},
  title   = {A Unifying Action Principle for Classical Mechanical Systems},
  journal = {Physical Review E},
  year    = {2024},
  doi     = {10.1103/rch9-wsbw},
  url     = {https://journals.aps.org/pre/abstract/10.1103/rch9-wsbw},
  eprint = "2409.11063",
  archivePrefix = "arXiv",
  primaryClass = "physics.class-ph",
}

@article{de_Le_n_2024,
   title={A new perspective on nonholonomic brackets and Hamilton–Jacobi theory},
   volume={198},
   ISSN={0393-0440},
   url={http://dx.doi.org/10.1016/j.geomphys.2024.105116},
   DOI={10.1016/j.geomphys.2024.105116},
   journal={Journal of Geometry and Physics},
   publisher={Elsevier BV},
   author={Manuel de Le\'on and Manuel Lainz and Asier L\'opez-Gord\'on and Juan Carlos Marrero},
   year={2024},
   month=apr, pages={105116} }

@article{FernandezRadhakrishnan2018,
  author  = {Oscar E. Fernandez and Mala L. Radhakrishnan},
  title   = {The Quantum Mechanics of a Rolling Molecular Nanocar},
  journal = {Scientific Reports},
  volume  = {8},
  pages   = {14878},
  year    = {2018},
  doi     = {10.1038/s41598-018-33023-8}
}

@article{Horowitz:2024eea,
    author = "Horowitz, W. A. and Rothkopf, A.",
    title = "{Even More Generalized Hamiltonian Dynamics}",
    eprint = "2408.14420",
    archivePrefix = "arXiv",
    primaryClass = "math-ph",
    month = "8",
    year = "2024"
}

@book{GelfandFomin2000,
  author    = {I. M. Gelfand and S. V. Fomin},
  title     = {Calculus of Variations},
  publisher = {Dover},
  address   = {Mineola, NY},
  year      = {2000}
}

@book{Popper1959,
  author    = {Karl R. Popper},
  title     = {The Logic of Scientific Discovery},
  publisher = {Routledge},
  address   = {London},
  year      = {1959}
}

@article{Einstein1905,
  author  = {Albert Einstein},
  title   = {Zur Elektrodynamik bewegter K{\"o}rper},
  journal = {Annalen der Physik},
  volume  = {17},
  pages   = {891--921},
  year    = {1905},
  doi     = {10.1002/andp.19053221004}
}

@article{Hamilton1834,
  author  = {William Rowan Hamilton},
  title   = {On a General Method in Dynamics},
  journal = {Philosophical Transactions of the Royal Society of London},
  volume  = {124},
  pages   = {247--308},
  year    = {1834}
}

@article{Hamilton1835,
  author  = {William Rowan Hamilton},
  title   = {Second Essay on a General Method in Dynamics},
  journal = {Philosophical Transactions of the Royal Society of London},
  volume  = {125},
  pages   = {95--144},
  year    = {1835}
}

@book{Newton1687,
  author    = {Isaac Newton},
  title     = {Philosophi{\ae} Naturalis Principia Mathematica},
  year      = {1687},
  publisher = {Royal Society},
  address   = {London}
}

@misc{Rothkopf2026,
  author       = {Alexander Rothkopf and W. A. Horowitz and J. Nordstr{\"o}m},
  howpublished = {In preparation},
  year         = {2026}
}

@book{Wong2001,
  author    = {Roderick Wong},
  title     = {Asymptotic Approximations of Integrals},
  publisher = {Society for Industrial and Applied Mathematics (SIAM)},
  address   = {Philadelphia, PA},
  year      = {2001},
  series    = {Classics in Applied Mathematics},
  volume    = {34},
  isbn      = {978-0-89871-926-0},
  doi       = {10.1137/1.9780898719260},
  url       = {https://doi.org/10.1137/1.9780898719260},
  note      = {Corrected reprint of the 1989 edition}
}

@article{DeWitt:1957at,
    author = "DeWitt, Bryce S.",
    title = "{Dynamical theory in curved spaces. 1. A Review of the classical and quantum action principles}",
    doi = "10.1103/RevModPhys.29.377",
    journal = "Rev. Mod. Phys.",
    volume = "29",
    pages = "377--397",
    year = "1957"
}

@article{Trotter:1959ytf,
    author = "Trotter, H. F.",
    title = "{On the product of semi-groups of operators}",
    doi = "10.1090/s0002-9939-1959-0108732-6",
    journal = "Proc. Am. Math. Soc.",
    volume = "10",
    number = "4",
    pages = "545--551",
    year = "1959"
}

@book{Wolff2003,
  author       = {Wolff, Thomas H.},
  editor       = {Łaba, Izabella and Shubin, Carol},
  title        = {Lectures on Harmonic Analysis},
  series       = {University Lecture Series},
  volume       = {29},
  publisher    = {American Mathematical Society},
  address      = {Providence, RI},
  year         = {2003},
  isbn         = {978-0-8218-3449-7},
  ebookisbn    = {978-1-4704-1837-3},
  pages        = {137},
  url          = {https://www.ams.org/bookpages/ulect-29}
}

@book{CoddingtonLevinson1955,
  author    = {Coddington, Earl A. and Levinson, Norman},
  title     = {Theory of Ordinary Differential Equations},
  publisher = {McGraw-Hill},
  year      = {1955},
  note      = {9th reprint (1987)},
  isbn      = {9780070992566}
}

@article{Berges:2007ym,
    author = "Berges, Juergen and Gasenzer, Thomas",
    title = "{Quantum versus classical statistical dynamics of an ultracold Bose gas}",
    eprint = "cond-mat/0703163",
    archivePrefix = "arXiv",
    reportNumber = "HD-THEP-07-05",
    doi = "10.1103/PhysRevA.76.033604",
    journal = "Phys. Rev. A",
    volume = "76",
    pages = "033604",
    year = "2007"
}

@article{Kugo:1979gm,
    author = "Kugo, Taichiro and Ojima, Izumi",
    title = "{Local Covariant Operator Formalism of Nonabelian Gauge Theories and Quark Confinement Problem}",
    reportNumber = "KUNS-493",
    doi = "10.1143/PTPS.66.1",
    journal = "Prog. Theor. Phys. Suppl.",
    volume = "66",
    pages = "1--130",
    year = "1979"
}

@article{Strang1968,
  author  = {Strang, Gilbert},
  title   = {On the Construction and Comparison of Difference Schemes},
  journal = {SIAM Journal on Numerical Analysis},
  volume  = {5},
  number  = {3},
  pages   = {506--517},
  year    = {1968},
  doi     = {10.1137/0705041}
}

@article{Schwinger1961,
  author  = {Julian Schwinger},
  title   = {Brownian Motion of a Quantum Oscillator},
  journal = {Journal of Mathematical Physics},
  volume  = {2},
  number  = {3},
  pages   = {407--432},
  year    = {1961},
  doi     = {10.1063/1.1703727}
}

@article{Keldysh1965,
  author  = {L. V. Keldysh},
  title   = {Diagram Technique for Nonequilibrium Processes},
  journal = {Soviet Physics JETP},
  volume  = {20},
  pages   = {1018--1026},
  year    = {1965}
}

@book{Laine:2016hma,
    author = "Laine, Mikko and Vuorinen, Aleksi",
    title = "{Basics of Thermal Field Theory}",
    eprint = "1701.01554",
    archivePrefix = "arXiv",
    primaryClass = "hep-ph",
    doi = "10.1007/978-3-319-31933-9",
    publisher = "Springer",
    volume = "925",
    year = "2016"
}

@article{Jordan1986,
  author  = {Richard D. Jordan},
  title   = {Effective Field Equations for Expectation Values},
  journal = {Physical Review D},
  volume  = {33},
  number  = {2},
  pages   = {444--454},
  year    = {1986},
  doi     = {10.1103/PhysRevD.33.444}
}

@article{Weinberg2005,
  author  = {Steven Weinberg},
  title   = {Quantum Contributions to Cosmological Correlations},
  journal = {Physical Review D},
  volume  = {72},
  pages   = {043514},
  year    = {2005},
  doi     = {10.1103/PhysRevD.72.043514}
}

@book{Kamenev2011,
  author    = {Alex Kamenev},
  title     = {Field Theory of Non-Equilibrium Systems},
  publisher = {Cambridge University Press},
  address   = {Cambridge},
  year      = {2011},
  doi       = {10.1017/CBO9781139003666}
}

@book{Casalderrey-Solana:2011dxg,
    author = "Casalderrey-Solana, Jorge and Liu, Hong and Mateos, David and Rajagopal, Krishna and Achim Wiedemann, Urs",
    title = "{Gauge/String Duality, Hot QCD and Heavy Ion Collisions}",
    eprint = "1101.0618",
    archivePrefix = "arXiv",
    primaryClass = "hep-th",
    reportNumber = "CERN-PH-TH-2010-316, MIT-CTP-4198, ICCUB-10-202",
    doi = "10.1017/9781009403504",
    isbn = "978-1-009-40350-4, 978-1-009-40349-8, 978-1-009-40352-8, 978-1-139-13674-7",
    publisher = "Cambridge University Press",
    year = "2014"
}

@book{Das:1997gg,
    author = "Das, Ashok K.",
    title = "{Finite Temperature Field Theory}",
    isbn = "978-981-02-2856-9, 978-981-4498-23-4",
    publisher = "World Scientific",
    address = "New York",
    year = "1997"
}

@book{Rammer2007,
  author    = {J{\o}rgen Rammer},
  title     = {Quantum Field Theory of Non-equilibrium States},
  publisher = {Cambridge University Press},
  address   = {Cambridge},
  year      = {2007},
  isbn      = {978-0-521-87499-4},
  doi       = {10.1017/CBO9780511619433}
}

@article{Berges:2004yj,
    author = "Berges, Juergen",
    editor = "Bracco, Mirian and Chiapparini, Marcelo and Ferreira, Erasmo and Kodama, Takeshi",
    title = "{Introduction to nonequilibrium quantum field theory}",
    eprint = "hep-ph/0409233",
    archivePrefix = "arXiv",
    doi = "10.1063/1.1843591",
    journal = "AIP Conf. Proc.",
    volume = "739",
    number = "1",
    pages = "3--62",
    year = "2004"
}

@book{BalzerBonitz2013,
  author    = {Karsten Balzer and Michael Bonitz},
  title     = {Nonequilibrium Green's Functions Approach to Inhomogeneous Systems},
  publisher = {Springer},
  address   = {Berlin, Heidelberg},
  year      = {2013},
  series    = {Lecture Notes in Physics},
  volume    = {867},
  doi       = {10.1007/978-3-642-35085-9},
  isbn      = {978-3-642-35084-2}
}

@article{Flannery2005_enigma_nonholonomic,
  title        = {The enigma of nonholonomic constraints},
  author       = {Flannery, M. R.},
  journal      = {American Journal of Physics},
  volume       = {73},
  number       = {3},
  pages        = {265--272},
  year         = {2005},
  publisher    = {American Association of Physics Teachers},
  doi          = {10.1119/1.1830501}
}

@article{Bert:2025ffu,
    author = "Bert, Ben and Horowitz, William A.",
    title = "{Integrable Non-Holonomic Constraints and Gauge Fixing in Classical Field Theory}",
    eprint = "2505.20684",
    archivePrefix = "arXiv",
    primaryClass = "hep-th",
    month = "5",
    year = "2025",
}

@book{Kleinert2009,
  author    = {Hagen Kleinert},
  title     = {Path Integrals in Quantum Mechanics, Statistics, Polymer Physics, and Financial Markets},
  publisher = {World Scientific},
  address   = {Singapore},
  year      = {2009},
  doi       = {10.1142/7305},
  isbn      = {9789814365260}
}

@book{FeynmanHibbs1965,
  author    = {Richard P. Feynman and A. R. Hibbs},
  title     = {Quantum Mechanics and Path Integrals},
  publisher = {McGraw--Hill},
  address   = {New York},
  year      = {1965}
}

@article{Marinov:1980vn,
    author = "Marinov, M. S.",
    title = "{PATH INTEGRALS IN QUANTUM THEORY: AN OUTLOOK OF BASIC CONCEPTS}",
    doi = "10.1016/0370-1573(80)90111-8",
    journal = "Phys. Rept.",
    volume = "60",
    pages = "1--57",
    year = "1980"
}

@book{Schulman:1981vu,
    author = "Schulman, L. s.",
    title = "{TECHNIQUES AND APPLICATIONS OF PATH INTEGRATION}",
    year = "1981"
}

@article{Rothkopf:2024hxi,
    author = {Rothkopf, Alexander and Horowitz, W. A. and Nordstr{\"o}m, Jan},
    title = "{Exact symmetry conservation and automatic mesh refinement in discrete initial boundary value problems}",
    eprint = "2404.18676",
    archivePrefix = "arXiv",
    primaryClass = "math.NA",
    doi = "10.1016/j.jcp.2024.113686",
    journal = "J. Comput. Phys.",
    volume = "524",
    pages = "113686",
    year = "2025"
}

@article{Rothkopf:2023ljz,
    author = {Rothkopf, Alexander and Nordstr{\"o}m, Jan},
    title = "{A symmetry and Noether charge preserving discretization of initial value problems}",
    eprint = "2307.04490",
    archivePrefix = "arXiv",
    primaryClass = "math.NA",
    doi = "10.1016/j.jcp.2023.112652",
    journal = "J. Comput. Phys.",
    volume = "498",
    pages = "112652",
    year = "2024"
}

@article{Dunne:2007rt,
    author = "Dunne, Gerald V.",
    editor = "Gadella, Manuel and Izquierdo, Jose M. and Kuru, Sengul and Negro, Javier and del Olmo, Mariano A.",
    title = "{Functional determinants in quantum field theory}",
    eprint = "0711.1178",
    archivePrefix = "arXiv",
    primaryClass = "hep-th",
    doi = "10.1088/1751-8113/41/30/304006",
    journal = "J. Phys. A",
    volume = "41",
    pages = "304006",
    year = "2008"
}

@book{brouwer2011spectra,
  title={Spectra of Graphs},
  author={Brouwer, Andries E. and Haemers, Willem H.},
  year={2012},
  publisher={Springer Science \& Business Media},
  series={Universitext},
  doi={10.1007/978-1-4614-1939-6},
  isbn={978-1-4614-1938-9}
}

@article{Flannery2011DAlembertLagrange,
  author  = {Flannery, M. R.},
  title   = {d{’}Alembert--Lagrange analytical dynamics for nonholonomic systems},
  journal = {Journal of Mathematical Physics},
  volume  = {52},
  number  = {3},
  pages   = {032705},
  year    = {2011},
  doi     = {10.1063/1.3559128},
}

@book{HornJohnsonMatrixAnalysis,
  author    = {Horn, Roger A. and Johnson, Charles R.},
  title     = {Matrix Analysis},
  edition   = {2},
  publisher = {Cambridge University Press},
  address   = {Cambridge},
  year      = {2013},
  isbn      = {978-0-521-83940-2},
}

@book{Hormander2003,
  author    = {Lars H{\"o}rmander},
  title     = {{The Analysis of Linear Partial Differential Operators I: Distribution Theory and Fourier Analysis}},
  series    = {Classics in Mathematics},
  edition   = {2},
  publisher = {Springer Berlin Heidelberg},
  year      = {2003},
  doi       = {10.1007/978-3-642-61497-2},
  isbn      = {978-3-642-61497-2},
  note      = {Softcover ISBN: 978-3-540-00662-6},
  url       = {https://doi.org/10.1007/978-3-642-61497-2}
}

@article{Galley:2012hx,
    author = "Galley, Chad R.",
    title = "{Classical Mechanics of Nonconservative Systems}",
    eprint = "1210.2745",
    archivePrefix = "arXiv",
    primaryClass = "gr-qc",
    doi = "10.1103/PhysRevLett.110.174301",
    journal = "Phys. Rev. Lett.",
    volume = "110",
    number = "17",
    pages = "174301",
    year = "2013"
}

@book{goldstein2002classical,
  author    = {Herbert Goldstein and Charles Poole and John Safko},
  title     = {Classical Mechanics},
  edition   = {3rd},
  publisher = {Addison Wesley},
  year      = {2002},
  isbn      = {9780201657029}
}
\bibliographystyle{apsrev4-1}

\end{document}